\documentclass[aps,prl,twocolumn,superscriptaddress]{revtex4-2}

\usepackage{amsmath}
\usepackage{graphicx}
\usepackage{dcolumn}
\usepackage{bm}
\usepackage{braket}
\usepackage{xcolor}
\usepackage[normalem]{ulem} 

\newcommand{\beginsupplement}{%
	\setcounter{table}{0}
	\renewcommand{\thetable}{S\arabic{table}}%
	\setcounter{figure}{0}
	\renewcommand{\thefigure}{S\arabic{figure}}%
}

\begin{document}
	
	
	\title{Readout-Induced Suppression and Enhancement of Superconducting Qubit 
		Lifetimes}

	\author{Ted Thorbeck}
	\email{ted.thorbeck@ibm.com}
	\affiliation{IBM Quantum, IBM T.J. Watson Research Center, Yorktown Heights, NY 10598, USA}
	\author{Zhihao Xiao}
	\affiliation{Department of Physics and Applied Physics, University of Massachusetts, Lowell, MA 01854, USA}
	\author{Archana Kamal}
	\affiliation{Department of Physics and Applied Physics, University of Massachusetts, Lowell, MA 01854, USA}
	\author{Luke C. G. Govia}
	\affiliation{IBM Quantum, IBM Almaden Research Center, San Jose, CA 95120, USA}
	
	\date{\today}
	
	\begin{abstract}
		It has long been known that the lifetimes of superconducting qubits suffer during readout, increasing readout errors. We show that this degradation is due to the anti-Zeno effect, as readout-induced dephasing broadens the qubit so that it overlaps `hot spots’ of strong dissipation, likely due to two-level systems in the qubit's bath. Using a flux-tunable qubit to probe the qubit's frequency dependent loss, we accurately predict the change in lifetime during readout with a new self-consistent master equation that incorporates the modification to qubit relaxation due to measurement-induced dephasing. Moreover, we controllably demonstrate both the Zeno and anti-Zeno effects, which explain suppression and the rarer enhancement of qubit lifetimes during readout.

	\end{abstract}
	
	\pacs{Valid PACS appear here}
	\maketitle

	\emph{Introduction--}The speed and simplicity of dispersive readout has made it the dominant readout technique for superconducting qubits, despite long-standing mysteries about why dispersive readout sometimes destroys the state of the qubit rather than faithfully reporting it \cite{blais2004cavity, blais2021circuit, wallraff2005approaching}. These are called non-QND (quantum nondemolition) errors, to distinguish them from the trivial errors that occur when the measurement tone is too weak, or too much readout signal is lost, and there is not enough information to distinguish the qubit states. Readout errors threaten quantum computing, because they add to the overhead for long-term goals like quantum error correction or near-term applications such as error mitigation \cite{bravyi2021mitigating, vandenberg2022model, fowler2012surface, leymann2020bitter}.  They also frustrate dynamic circuits, in which operations are conditioned on the outcome of mid-circuit measurements \cite{corcoles2021exploiting}.
	
	Relaxation and leakage are the two dominant non-QND errors during readout.
	Leakage is when the transmon leaves the qubit subspace, by, for example, occupying one of the higher levels of a transmon. Recent work has shown that readout-induced leakage can occur when a transition frequency from the qubit subspace to a highly excited state is resonant with an occupied resonator state, allowing excitations to swap into the transmon \cite{sank2016measurement, khezri2022measurement, shillito2022dynamics, cohen2022reminiscence, verney2019structural, lescanne2019escape, malekakhlagh2022optimization}. Relaxation during readout is expected because the qubit has a finite lifetime, $T_1$, but $T_1$ is often suppressed during readout, resulting in an excess of readout errors.  The suppression does not need to be monotonic as a function of measurement strength, and sometimes $T_1$ can even show an anomalous increase during readout \cite{harrington2017quantum, johnson2012heralded, minev2019catch, gusenkova2021quantum, sivak2022real, mallet2009single}. Early work showed that backaction from the measurement apparatus could degrade $T_1$, but this has been mitigated by adding isolation between the readout resonator and the amplifier \cite{picot2008role, serban2010relaxation, mallet2009single}. Another possible explanation is dressed dephasing \cite{boissonneault2008nonlinear, boissonneault2009dispersive, slichter2012measurement}, in which photons in the resonator combine with dephasing noise at the detuning frequency between the qubit and the readout resonator, causing excitation or relaxation in the qubit. However, this can be minimized by protecting the qubit from dephasing noise at the detuning frequency.  
	
	Much of the recent theoretical work on $T_1$ suppression during readout has focused on how the readout drive changes the Purcell loss \cite{sete2014purcell, malekakhlagh2020lifetime, petrescu2020lifetime, hanai2021intrinsic, boissonneault2010improved, sete2015quantum, muller2020dissipative}. Purcell loss is dissipation induced on the qubit due to its coupling to the lossy readout resonator. As the strength of the measurement tone changes the qubit-resonator hybridization, it also changes the Purcell loss. Initial models predicted only an increase in $T_1$, but more recent theoretical treatments predict parameter regimes where a decrease in $T_1$ is possible. This qualitatively, if not quantitatively, agrees with experiment. However Purcell loss is typically engineered to be very small via a Purcell filter \cite{reed2010fast, jeffrey2014fast}, so increasing it is unlikely to be the origin of excess relaxation.  Therefore the dominant cause of the suppression of $T_1$ during readout is still a mystery, and solving it will not only help us engineer protection into quantum processors, but also answer a long-standing, unresolved question in the field of open quantum systems.

	In this letter we show that the change in the lifetime of superconducting qubits during readout is due to the quantum Zeno and anti-Zeno effects. The quantum Zeno effect predicts that measurement should freeze a quantum system, preventing radiative decay and increasing the system’s lifetime \cite{misra1977zeno, itano1990quantum}. However, it was discovered that the opposite effect, measurement increasing the rate of decay, called the anti-Zeno effect, is far more common \cite{kofman2000acceleration, kofman2001frequent, kofman2001zeno}. The Zeno effect has been experimentally demonstrated in superconducting qubits \cite{kakuyanagi2015observation, harrington2017quantum, slichter2016quantum}, and it has been used for qubit control \cite{hacohen2018incoherent} and gates \cite{blumenthal2022demonstration}, but it has not been invoked to explain the change in $T_1$ during readout of superconducting qubits. 
	
	We begin by reinterpreting the anti-Zeno effect in terms familiar to the circuit-QED community: during readout the qubit's transition is both broadened by dephasing and 
	Stark-shifted. $T_1$ is typically reduced during readout because the broadened qubit is more likely to interact with `hot spots' of higher dissipation.  We believe that in our device TLS (two-level systems) are the dominant source of `hot spots', but any parasitic device or package modes would show similar behaviour.  We show this by using a flux-tunable qubit to map the qubit's decay rate as a function of frequency, from which we can then predict how $T_1$ changes during readout. While it has previously been suggested that readout can be degraded if the qubit is Stark-shift into a TLS \cite{sank2016measurement, sivak2022real}, we show that it is dephasing, not Stark shift, that is essential to understanding the suppression of qubit lifetimes during readout.  We also show that this is the dominant mechanism to explain the change in qubit lifetime during readout in a modern superconducting quantum processor. While the Zeno and anti-Zeno effects are well known, modeling these effects challenges the traditional Lindbladian master equation approach for dissipation in superconducting circuits, because the qubit interacts with two baths, one for dephasing and the other for dissipative decay.  During readout, the interaction with the dephasing bath is much stronger than the interaction with the dissipative bath, so our model needs to capture how the first bath influences the system-bath interaction of the second. Using the methodology of Ref.~\cite{DISCO} we show how to self-consistently address the impact of dephasing on the qubit's coupling to the dissipative bath, and extend the Kofman-Kurizki formula for the Zeno/Anti-Zeno effect to encompass coherent couplings to an ancilla system \cite{kofman2000acceleration, kofman2001frequent, kofman2001zeno}.

	\begin{figure}[t!]
		\includegraphics[]{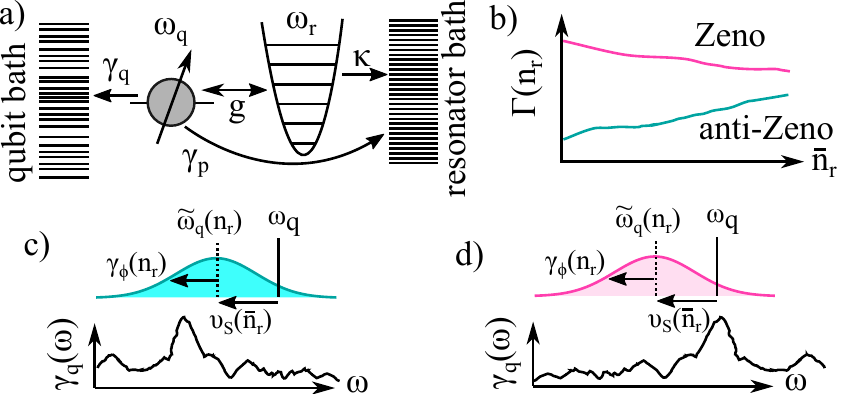}
		\caption{Fundamentals of Zeno and anti-Zeno effects. (a) In dispersive readout the qubit, $\omega_q$, is coupled to the readout resonator, $\omega_r$, via the Jaynes-Cummings Hamiltonian, with coupling strength $g$. The qubit and resonator are each coupled to their own continuum of bath modes into which they can irreversibly lose an excitation. The qubit's (resonator's) dissipation rate to it's own bath is $\gamma_q$ ($\kappa$). Because the qubit is coupled to the readout resonator, the qubit can also dissipate into the resonator's bath (Purcell loss) at a rate $\gamma_P$. The measured qubit lifetime is $1/T_1 = \gamma_q + \gamma_P$. In this device $\gamma_P \ll \gamma_q$ because of a Purcell filter.
			(b) Schematic of how the decay rate during readout, $\Gamma(\bar{n}_r)$ depends on the readout strength, in terms of the average number of photons in the resonator, for both the Zeno and anit-Zeno effects.
			(c)  The qubit decay rate, in the absence of readout, is dominated by a few hot spots, likely due to parasitic TLS.  During readout the qubit's frequency is Stark shifted by $\nu_S(\bar{n}_r)$ and broadened at the dephasing rate $\gamma_\phi(\bar{n}_r)$, as shown by the Lorentzian centered about the Stark shifted qubit frequency, $\tilde{\omega}_q(\bar{n}_r)$. Here the Lorentzian overlaps a hot spot that the qubit is not sensitive to its natural frequency, thus increasing the decay rate.
			(d) To display the Zeno effect, the natural qubit frequency must be resonant with a hot spot, so that during readout the qubit becomes less sensitive to the hot spot. } 
		\label{basic_idea}
	\end{figure}

	\emph{Anti-Zeno Effect and $T_1$ During Readout--}We begin by explaining the anti-Zeno effect in circuit-QED terminology, Fig. 1(a). From qubit lifetime spectroscopy, measuring $T_1$ as a function of the qubit frequency, it is well known that $\gamma_q = 1/T_1$ is not a constant; there are hot spots, likely due to TLS or packaging modes, where the decay rate is higher (Fig. 1(c)) \cite{klimov2018fluctuations, carroll2022dynamics, barends2013coherent, lisenfeld2023enhancing, muller2019towards}. To understand how the qubit lifetime changes during measurement, the Zeno and anti-Zeno effects, we neglect the dynamics of the readout resonator to focus only on how the measurement affects the qubit.  First, during readout the qubit frequency is Stark shifted, leading to a detuning from its natural frequency by $\nu_S = 2 \chi \bar{n}_r$, where $\chi$ is the dispersive shift and $\bar{n}_r$ is the average number of photons in the readout resonator. Second, measurement destroys the qubit's coherence, i.e.~it dephases the qubit, as described by an exponential decay rate, $\gamma_\phi(\bar{n}_r)$.  Dephasing can be understood as uncertainty in the qubit’s frequency, so during readout we need to consider all possible qubit frequencies instead of solely considering the natural qubit frequency. As derived in Ref. \cite{supplement, DISCO}, the Kofman–Kurizki formula \cite{kofman2000acceleration, kofman2001frequent, kofman2001zeno} relates the qubit decay rate during continuous measurement,
	\begin{equation}
		\Gamma(\bar{n}_r) = \int_{-\infty}^{\infty} \gamma_q(\omega) \frac{1}{\pi}\frac{\gamma_\phi(\bar{n}_r)}{\gamma_\phi(\bar{n}_r)^2 + \left(\omega - \tilde{\omega}_q(\bar{n}_r)   \right)^2} \,d\omega ,
	\end{equation}
	to an integral that computes the average of the qubit's decay rate over frequency weighted by a Lorentzian centered on the Stark-shifted qubit frequency, $\tilde{\omega}_q(\bar{n}_r) = \omega_q + \nu_S$, with a width equal to the dephasing rate. This Lorentzian is similar to the qubit absorption spectrum during readout derived in \cite{gambetta2006qubit, supplement}. As the measurement strength goes to zero, $\bar{n}_r, \nu_S, \gamma_\phi \rightarrow 0$, the Lorentzian approaches a delta function, so the decay rate reduces to that given by Fermi's golden rule at the natural qubit frequency, $\Gamma \rightarrow \gamma_q(\omega_q)$ \cite{supplement}. 
	
	The key insight from the anti-Zeno effect literature is that evaluating the Kofman-Kurizki formula is more likely to result in the anti-Zeno effect than the Zeno effect  \cite{kofman2000acceleration, kofman2001frequent, kofman2001zeno}. Consider a typical plot of $\gamma_q(\omega)$, Fig. 1(c), which has a hot spot below the natural qubit frequency. The Lorentzian shows that as readout strength is increased, the qubit is Stark shifted to lower frequencies and broadened by the dephasing, until the Lorentzian overlaps this hot spot. As shown in Fig. 1(b) the Kofman-Kurizki formula predicts an increase in the decay rate: the anti-Zeno effect. The Zeno effect can be observed when the natural qubit frequency is aligned with a hot spot, so that measurement moves the qubit to a region of slower decay, Fig. 1(d). Because the qubit is not typically resonant with a hot spot, the anti-Zeno effect is more common than the Zeno effect.  
	
	\begin{figure}[t!]
		\includegraphics[]{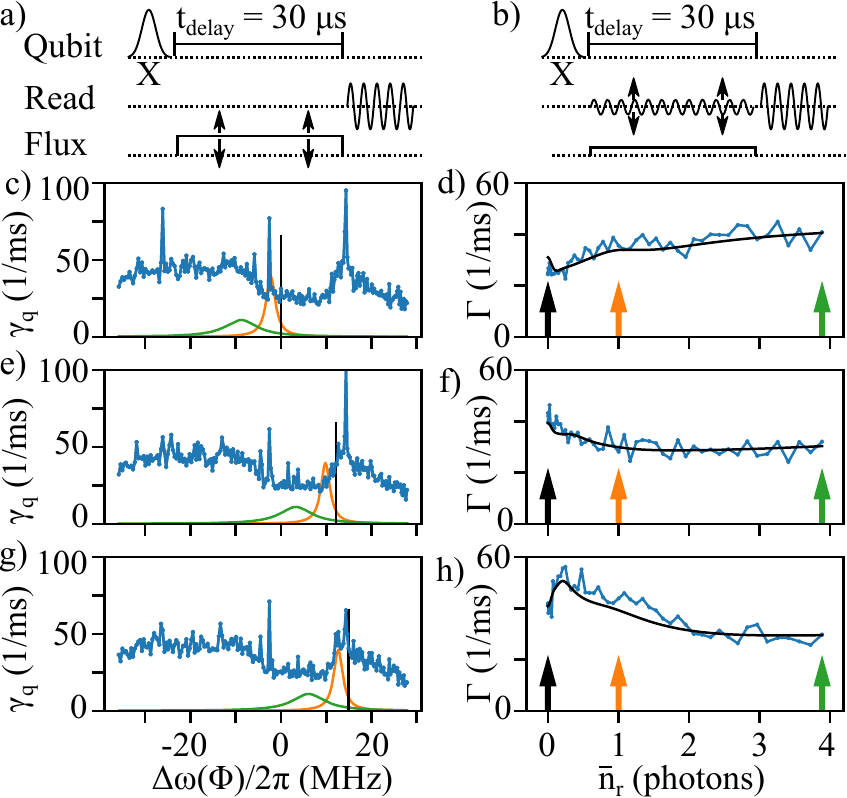}
		\caption{Predicting $\Gamma(\bar{n}_r)$ using the Kofman-Kurizki formula. (a) Fixed-delay $T_1$ experiment, with $t_{delay}$ = 30~$\mu$s, to measure the qubit decay rate $\gamma_q(\omega)$ as a function of frequency.  A flux pulse tuned the qubit frequency by $\Delta\omega(\Phi)$ about the natural qubit frequency, $\omega_q/2\pi$= 4.884~GHZ. We convert qubit population at the end of the experiment, $p_1$, to a decay rate using $\gamma_q = -\log(p_1)/t_{delay}$, neglecting heating and imperfect initialization. (b) Measurement of decay rate during readout, $\Gamma(\bar{n}_r)$, as a function of strength of the psuedo-measurement pulse, the output of which is not recorded. We interleave these two sweeps into a single experiment, lasting about 100~s, to prevent drift in $\gamma_q(\omega)$. (c, e, g) Experimental results of $\gamma_q(\omega)$ sweep in blue. (d, f, h) Experimental results of $\Gamma(\bar{n}_r)$ sweep in blue, with the Kofman-Kurizki prediction in black. A flux offset is applied during psuedo-measurement tone to shift the portion of $\gamma_q(\omega)$ that is sampled during readout, demonstrating the ability to switch between the Zeno and anti-Zeno effects. The flux-tuned qubit frequency at $\bar{n}_r$ = 0, black arrow on right, is indicated by the vertical black line on the left. The colored Lorentzians show the portions of $\gamma_q(\omega)$ involved in evaluating the Kofman-Kurizki formula at at $\bar{n}_r \approx$ 1 and 4, as shown by the colored arrows on the right.  
		} 
		\label{AZE_sweep}
	\end{figure}
	
	To test this theory, we compare the measured qubit decay rate during readout to the prediction from the Kofman–Kurizki formula. To measure $\gamma_q(\omega)$, we used a flux pulse to tune the qubit frequency during a fixed-delay $T_1$ experiment, Fig. 2(a). Then we measured the decay rate during readout, $\Gamma(\bar{n}_r)$, as a function of readout power, Fig. 2(b). The blue traces in Figs. 2(c-d) shows the results of these two measurements. Figure 2(d) shows $\Gamma(\bar{n}_r)$ increasing with readout power: the anti-Zeno regime. A numerical evaluation of the Kofman-Kurizki formula, black line in Fig. 2(d), successfully predicts the measured $\Gamma(\bar{n}_r)$. We can understand this prediction by looking at Fig. 2(c).  The natural qubit frequency, black line in Fig. 2(c) and black arrow in Fig. 2(d), is not aligned with any hot spots. As shown by the orange and green Lorentzians in Fig. 2(c), corresponding to the orange and green arrows in Fig. 2(d), the qubit becomes sensitive to a region with a higher qubit decay rate as the readout power increases.  Thus the qubit experiences a higher decay during readout.  
	
	By flux-tuning the qubit towards a hot spot, we can change the behaviour from anti-Zeno to Zeno in a predictable manner. In Fig. 2(c) there is a broad peak in the decay rate above the natural qubit frequency. To study this peak we add a flux pulse during the pseudo-measurement tone to position the qubit on the left shoulder of the peak during the measurement of $\Gamma(\bar{n}_r)$ as shown in Fig. 2(e-f).  Fig. 2(e) shows that the Stark shift and dephasing bring the qubit away from the peak and into a valley where the qubit decay rate is slower, which manifests in Fig. 2(f) as the decay rate decreasing with increasing readout strength, i.e.~the Zeno effect.  Next, in Fig. 2(g-h) we situate the qubit just to the right of the peak.  In Fig. 2(h) $\Gamma(\bar{n}_r)$ is non-monotonic, at first increasing and then decreasing. We interpret this in Fig. 2(g) as the readout at first bringing the qubit to the peak and then past it.  Thus the Kofman-Kurizki formula accurately quantifies the change in qubit $T_1$ observed during measurement in both the Zeno and anti-Zeno regimes.

	\begin{figure}[t!]
		\includegraphics[]{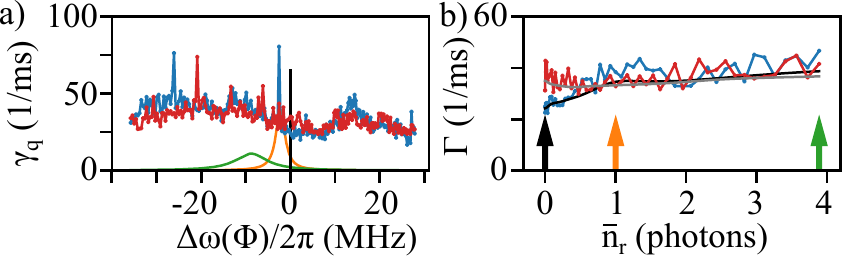}
		\caption{Changes in $\gamma_q(\omega)$ change the decay rate during readout.  The blue and the red curves show repetitions of the experiment from Fig. 2, taken 9 minutes apart.  The jump in (a) show up as changes in $\Gamma(\bar{n}_r$) in (b), as expected from the Kofman-Kurizki formula. } 
		\label{time_dep}
	\end{figure}
	
	\emph{Bath  dynamics--}The radiative environment of superconducting qubits is not static. For example, a TLS can jump in frequency after a nearby thermal fluctuator flips or after the impact of ionizing radiation \cite{thorbeck2022tls, muller2019towards}. The Kofman-Kurizki formula predicts $\Gamma(\bar{n}_r)$ will change as $\gamma_q(\omega)$ changes. We repeat the experiment from Fig. 2 to watch for sudden changes in $\gamma_q(\omega)$. Figure 3 shows two measurements, taken 9~minutes apart, which show a small changes in $\gamma_q(\omega)$ that is reflected in $\Gamma(\bar{n}_r)$.  Notably, both before and after the change, the Kofman-Kurizki formula accurately predicts the measured $\Gamma(\bar{n}_r)$. We emphasize that the anti-Zeno effect is the only theory for the change in $T_1$ during readout that can explain the time-dependent fluctuation of $T_1$ during readout.

	\begin{figure}[t!]
		\includegraphics[]{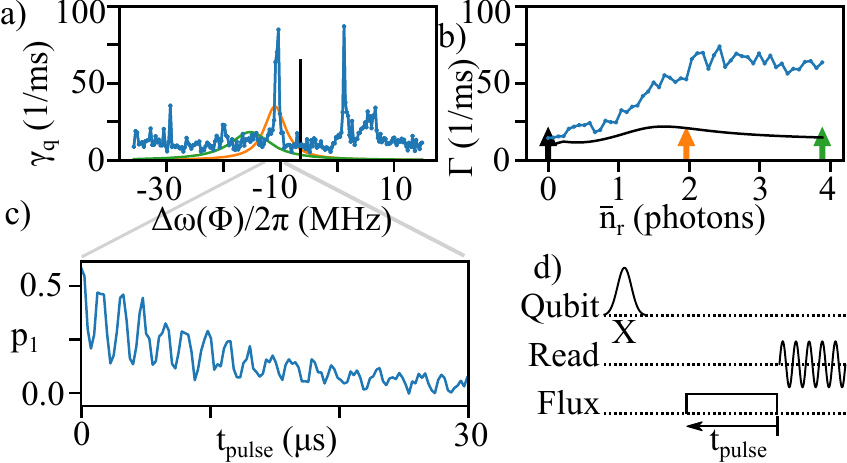}
		\caption{Breakdown of the Kofman-Kurizki prediction.  (a-b) Repetition of the experiment from Figs. 2 \& 3, taken about a month later, showing the failure Kofman-Kurizki formula to predict the measured $\Gamma(\bar{n}_r)$. Here the orange Lorentzians shows the dephasing broadened range of the qubit when the Stark shift is approximately equal to the detuning to the TLS, at about $\bar{n}_r \approx$ 2.
			(c) $p_1$ of the qubit during swap spectroscopy, showing oscillations between the qubit and the TLS.
			(d) Swap spectroscopy to look at the TLS at $\Delta/2\pi$ = -11 MHz in (a). The time between the $X$-pulse and the measurement tone was fixed at 50~$\mu$s, explaining why $p_1$ does not approach unity at $t_{\rm pulse}$ = 0 in (c). 
		} 
		\label{breakdown}
	\end{figure}

	\emph{Breakdown of Kofman-Kurizki--}We have shown that we can use the Kofman-Kurizki formula to predict the behaviour of $T_1$ during readout.
	However, describing a TLS as a mode in the bath spectrum is a coarse approximation that breaks down when a TLS coherently couples to the qubit. Figure 4 shows the Kofman-Kurizki prediction failing in this regime. The experimental conditions in Fig. 4 are identical to those in Figs. 2 \& 3, with the only change to the setup being the uncontrollable emergence of a strongly-coupled TLS at $\Delta\omega(\Phi)/2\pi$ = -11~MHz, reinforcing our claim that changes in $\gamma_q(\omega)$ cause the change in $\Gamma(\bar{n}_r)$. The data in Fig. 4(b) also clearly shows that the Stark shift alone cannot explain our results, given the absence of a narrow peak when the Stark shift equals the qubit-TLS detuning. 
	
	Figure 4(b) shows that the Kofman-Kurizki formula significantly underestimates the measured $\Gamma(\bar{n}_r)$. Swap spectroscopy, Fig. 4(c-d), shows vacuum-Rabi oscillations between the qubit and the TLS, breaking our analysis. After $t_{\rm pulse} = 30$~$\mu$s (used in Fig. 4(a-b)), the population is still oscillating, so fitting a single exponential is inappropriate. Moreover, the derivation of the Kofman-Kurizki formula makes the Markov assumption, that once the qubit decays to the bath the excitation cannot return \cite{supplement, DISCO, kofman2000acceleration, ai2013quantum}. The vacuum-Rabi oscillations in Fig. 4(c) demonstrate that the Markov approximation, and thus the Kofman-Kurizki formula, is no longer valid, likely because the qubit-TLS interaction is stronger than the bath equilibration rate (TLS intrinsic dissipation).

	\begin{figure}[t!]
		\includegraphics[]{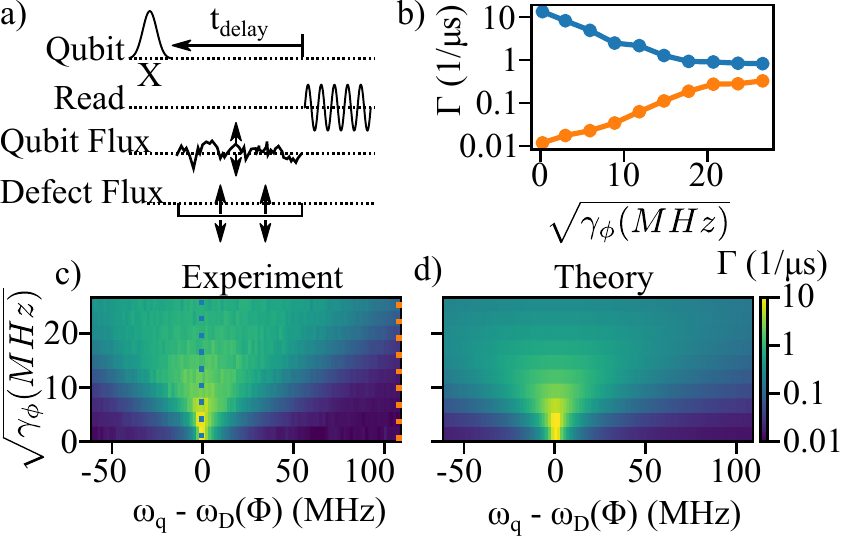}
		\caption{
			Zeno and anti-Zeno effects for a qubit coupled to a coherent defect. (a) Variable-delay $T_1$ experiment, with logarithmic spacing to reliably fit $\Gamma$ over orders of magnitude, as a function of both the magnitude of the flux noise applied to the qubit and the qubit-defect detuning.  The qubit frequency is fixed at $\omega_q/2\pi$ = 4.291~GHz, while the flux through the detect is swept. (b) Two line-cuts through (c): blue for qubit on resonance with the defect, showing the Zeno effect, and orange for the qubit far from resonance with the defect, showing the anti-Zeno effect. (c) Measured $\Gamma$ as a function of dephasing rate and qubit-defect detuning.  (d) Theory prediction of $\Gamma$ using Eq. 2 and parameters independently measured in  Ref. \cite{supplement}.
		} 
		\label{basic_idea}
	\end{figure}
	
	\emph{Artificial TLS--}Even in the case that the qubit interacts coherently with a single TLS, qubit relaxation can still be understood in terms of the Zeno and anti-Zeno effects. Given their uncontrollable dynamics, studying a natural TLS is challenging, so instead we manufacture a coherent TLS, that we will call the defect, by engineering a transmon with strong coupling to off-chip dissipation to have a very short lifetime, $\gamma_{1,D}$ = 1/(103~ns). It is coupled to the first transmon with strength $g_D/2\pi$ = 1.6~MHz \cite{supplement}. The defect is a quantum object, with its own bath, through which the qubit can decay via an effect analogous to Purcell decay. Our goal is to understand how qubit measurement impacts the decay of the qubit through the defect.
	
	Flux-tuning the qubit such that it is near resonant with the defect at $\omega_D/2\pi \approx$ 4.3~GHz brings it into a regime where it is very sensitive to flux noise. Thus, rather than using a pseudo-measurement tone, we add flux-noise to obtain enhanced qubit dephasing without a Stark-shift. As demonstrated in Ref. \cite{harrington2017quantum, alvarez2010zeno}, pure dephasing is a `quasi-measurement' that can cause the Zeno and anti-Zeno effects. One can also interpret our construction of the Kofman-Kurizki formula as highlighting that dephasing is the necessary ingredient for the Zeno/anti-Zeno effect, and the role of measurement is to supply that dephasing.  This highlights that  dephasing, not the Stark shift, is essential to explain changes in $T_1$ during readout.  
	
	Figure 5(c) shows the measured $\Gamma$ of the qubit as a function of qubit-defect detuning and flux-noise induced dephasing rate. For large detuning and small amounts of flux noise, $\Gamma \approx$ 1/100~$\mu$s is limited by the background $T_1$ of the qubit.  As the qubit comes into resonance with the defect, $\Gamma$ increases by three orders of magnitude. Line cuts through Fig. 5(c), shown in Fig. 5(b), demonstrate that our intuition for the regimes of the Zeno and anti-Zeno effects applies even for a strongly-coupled coherent defect \cite{cao2012transition}. Near the defect, adding dephasing noise decreases the decay rate (the Zeno effect), while away from the defect, adding noise increases the decay rate (the anti-Zeno effect). To capture this intuition, following Ref.~\cite{DISCO} and modeling the defect transmon as a harmonic oscillator we use adiabatic elimination to remove the defect and calculate the decay rate of the qubit into the defect's bath in the presence of dephasing on the qubit.  The total decay rate of the qubit, $\Gamma$, is then
	\begin{equation}
		\Gamma= \gamma_q +2 g_D^2 \frac{ \gamma_\phi + \gamma_{1,D}/2 - \gamma_q/2}{\left(\gamma_\phi + \gamma_{1,D}/2 - \gamma_q/2 \right)^2 + \left(\omega_q - \omega_D\right)^2}.
	\end{equation}
	If we ignored the loss ($\gamma_q$) and dephasing ($\gamma_\phi$) in the qubit, then we recover the standard formula for Purcell loss, so this equation is a generalization of Purcell loss that accounts for dephasing \cite{auffeves2010controlling}.  In Fig. 5(d) we plot $\Gamma$, showing good agreement with the experiment. All the parameters are independently measured in Ref. \cite{supplement}; there are no fitting parameters.  
	
	\emph{Conclusion--}We have shown that the quantum Zeno and anti-Zeno effects are the dominant cause of the suppression and enhancement of the decay rate of during readout for a modern superconducting qubit. We have successfully predicted how the decay rate changes during readout by accounting for how the measurement-induced Stark shift and dephasing change the qubit’s interaction with its radiative bath. This is a fundamental effect because, while a flux pulse can be used to counteract the Stark shift, the dephasing-induced broadening of the qubit is intrinsic to measurement. By changing the portion of $\gamma_q(\omega)$ that is sampled during readout, we demonstrated that it is possible to switch between the Zeno and anti-Zeno effects, which can be used to minimize the degradation of $T_1$ during readout.  Qubit tunability is routinely used to optimize the lifetime of the qubit \cite{lisenfeld2023enhancing, carroll2022dynamics, klimov2018fluctuations}, but to preserve the qubit lifetime during readout and to ensure truly QND readout, the optimization should consider the dephasing-broadened range of qubit frequencies during readout, rather than optimizing for the natural qubit frequency alone. Moreover, the anti-Zeno effect is an example of two decoherence mechanism interacting; the dephasing from readout changes the radiative dissipation. Our work represents experimental validation of the methodology of the self-consistent master equation \cite{DISCO}, lending confidence that similar problems can be understood and mitigated in other quantum systems using this technique.

	\textit{Acknowledgements} We thank Hannah Varekamp, Moein Malekakhlagh, Sami Rosenblatt, Hasan Nayfeh, Dave McKay, Muir Kumph, and Juzar Thinga for helpful conversations. The device used in this work was designed and fabricated internally at IBM Quantum. T.T. and L.G. were supported by the Army Research Office under QCISS (contract W911NF-21-1-0002) and device characterization by IARPA under LogiQ (W911NF-16-1-0114).  Z.X. and A.K. were supported by Air Force Office of Scientific Research (AFOSR) under grant FA9550-21-1-0151, and NSF under grant DMR-2047357.
	The views and conclusions contained in this document are those of the authors and should not be interpreted as representing the official policies, either expressed or implied, of the Army Research Office, IARPA, or the U.S. Government. The U.S. Government is authorized to reproduce and distribute reprints for Government purposes notwithstanding any copyright notation herein.

	\beginsupplement
	\renewcommand{\theequation}{S.\arabic{equation}}
	\setcounter{equation}{0}
	
	\clearpage
	
	\onecolumngrid
	\begin{center}
		\textbf{\large Supplementary Information for:\\ Readout-Induced Suppression and Enhancement of Superconducting Qubit Lifetimes}
	\end{center}
	\twocolumngrid
	
	\section{Calibration}
	\subsection{Stark shift and dephasing rate during measurement}
	
	\begin{figure}[h]
		\includegraphics[]{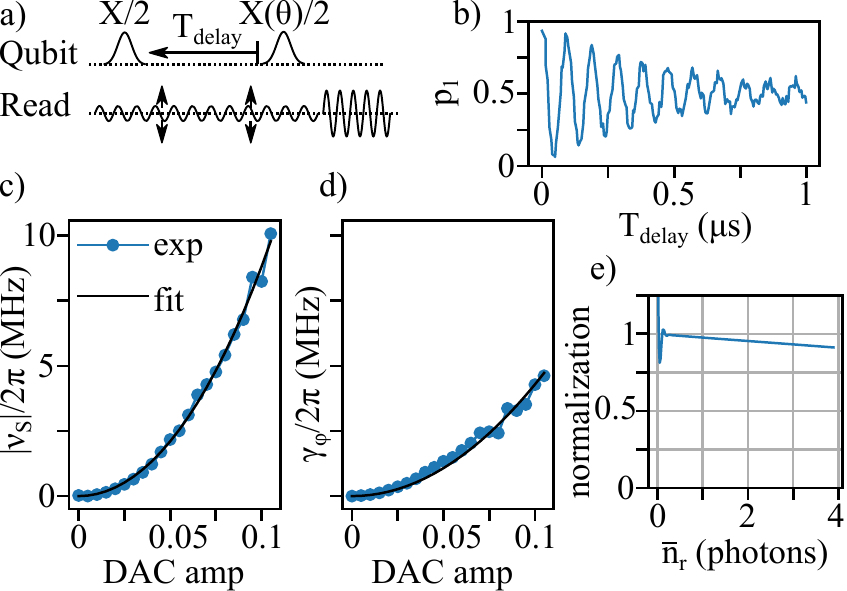}
		\caption{Calibration of Stark shift and dephasing rate during readout. (a) Ramsey experiment with driven readout resonator.  The duration of the pseudo-measurement tone is fixed at 5~$\mu$s, so the cavity is fully rung-up for the entire Ramsey experiment.  The psuedo-measurement tone ended 500~ns before the measurement pulse to let the cavity ring down.  A 10~MHz frequency offset to the Ramsey fringes is used to clearly distinguish the Stark shift from the dephasing rate.  This is added by rotating the angle, $\theta$, of the second pulse. (b) Ramsey fringes for a pseudo-measurement tone amplitude of 0.025 (DAC units) that we fit to a Stark shift, $\nu_S/2\pi$ = 0.451~MHz and dephasing rate, $\gamma_\phi/2\pi$ =  0.362~MHz.  (c-d) Plot of Stark shift (c) and dephasing rate (d) as a function of the amplitude of the pseudo-measurement tone.  (e) Normalization of the Lorentzian used when numerically evaluating the Kofman-Kurizki formula.} 
		\label{supp_photon_cals}
	\end{figure}
	
	Evaluating the Kofman-Kurizki formula requires knowing the Stark shift and the dephasing rate of the qubit as a function of the drive amplitude on the readout resonator. This is measured using a Ramsey fringe experiment while driving resonator as shown in Fig. \ref{supp_photon_cals}(a). By rotating the phase of the second pulse in the Ramsey experiment a 10~MHz frequency offset is added to the Ramsey fringes to clearly distinguish the time scales for the Stark shift and the dephasing rate. At each drive amplitude we fit the oscillations, Fig.~\ref{supp_photon_cals}(b), to a decaying sine wave, with decay rate $\gamma_\phi$ and frequency shift $\nu_S$. In Fig.~\ref{supp_photon_cals}(c) we fit the Stark shift as a function of strength of the readout drive $\epsilon$, in arbitrary DAC (digital-analog converter) units to a quadratic function with a small quartic correction for the Kerr effect,
	$\nu_S(\epsilon) = S \epsilon^2 + K \epsilon^4,$
	where $S/2\pi$ = 825~MHz and $K/2\pi$ = 5619~MHz.  The Kerr correction is small enough that it can be dropped. In Fig.~\ref{supp_photon_cals}(d) we fit the dephasing rate to a quadratic function
	$\gamma_\phi(\epsilon) = R \epsilon^2$,
	where $R / 2\pi$ = 429~MHz.  The Stark shift is then used to calibrate the number of photon in the readout resonator using $\nu_S = 2 \chi \bar{n}_r$.

	\subsection{Numerically evaluating Kofman-Kurizki integral }
	
	We numerically evaluate the Kofman-Kurizki formula using the \textit{numpy.trapz} function, introducing limitations at both the small drive limit and the large drive limit.  In the large drive limit, the integral is over $\omega$ from $-\infty$ to $\infty$, but our measurements of $\gamma_q(\omega)$ can only capture a finite frequency bandwidth. In the small drive limit, the dephasing rate approaches the natural dephasing rate, and the width of the Lorentzian is very small, smaller than the frequency step in our sampling of $\gamma_q(\omega)$.  We correct for both of these by dividing the raw data by a normalization given by integrating the Lorentzian alone, 
	\[ N = \int_{\omega_{low}}^{\omega_{high}} \frac{1}{\pi}\frac{\gamma_\phi(\bar{n}_r)}{\gamma_\phi(\bar{n}_r)^2 + \left(\omega - \tilde{\omega}_q(\bar{n}_r)   \right)^2} \,d\omega , \]
	where $\omega_{low}$ ($\omega_{high}$) is the lower (upper) bound of the measured frequency range.  As shown in Fig. \ref{supp_photon_cals}(e), aside from the first few points this is a very small correction.  This normalization can be interpreted as making the assumption that outside the measured frequency window, $\gamma_q(\omega)$ is well approximated by its average value within the measured window.
	
	\subsection{Calibrations for defect measurement}
	
	\begin{figure}[h]
		\includegraphics[]{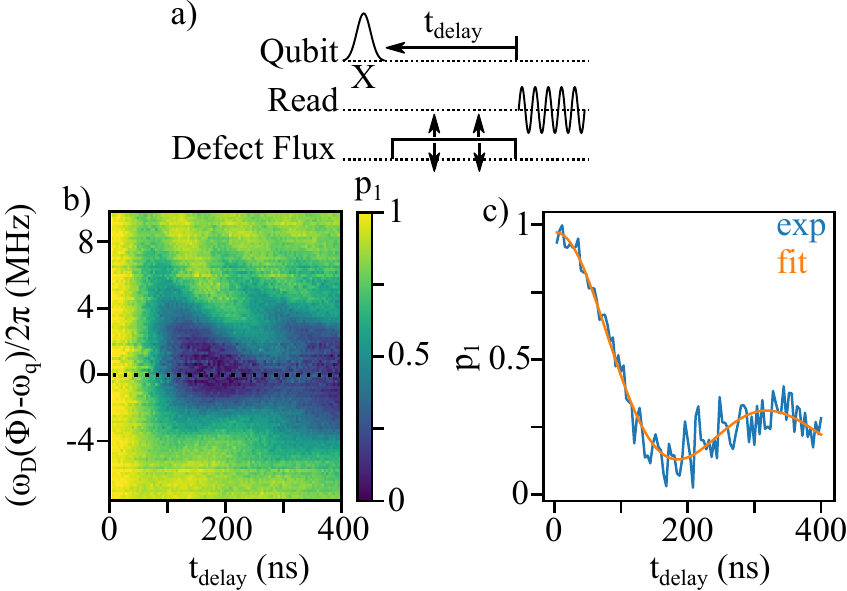}
		\caption{ (a) Ramsey experiment to measure coupling to and decay rate of the defect. 
			After an $X$-pulse prepares the qubit in the excited state, the duration and amplitude of a flux pulse are both swept.  
			(b) Excitation probability $p_1$ of the qubit at the end of the experiment, with a chevron pattern showing vacuum-rabi oscillations from the qubit to the defect and back. (c) A line-cut thru the blue dotted line in (b), where the amplitude of the oscillation is maximized, with a fit to a decaying sine function.
		}
		\label{supp_defect_cals}
	\end{figure}
	
	Figure \ref{supp_defect_cals}(a-c) shows the swap spectroscopy measurement of the coupling to the defect, $g_D$, and the decay rate of the defect, $\gamma_{1,D}$.  We fit the line-cut in Fig. \ref{supp_defect_cals}(b) to a damped sine function yielding $\gamma_{1,D}$ = 1/(103~ns).
	and $g_D/2\pi$ = 1.6~MHz.
	
\begin{figure}[h]
		\includegraphics[]{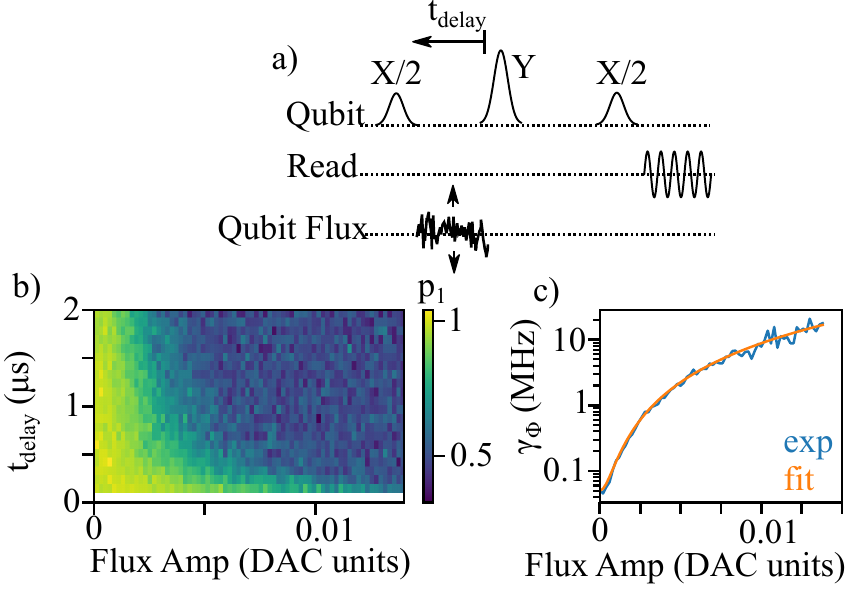}
		\caption{ Flux-noise calibration. (a) Ramsey echo experiment, with the flux noise applied only during the first free evolution period.  The echo removes any low-frequency background noise, so we only measure the applied flux noise. (b) Results of Ramsey echo, as a function of amplitude of the flux-noise signal.  We run the echo experiment for each amplitude of flux noise, with $t_{delay}$ swept from 0.1 to 2~$\mu$s. Each amplitude of flux noise is averaged over 50 randomization of the flux-noise signal. (c) The decay rates extracted from each echo sequence in (b) are fit to a quadratic function.
		} 
		\label{supp_flux_noise}
	\end{figure}

	We also convert the amplitude of the applied flux noise (in DAC units) to a dephasing rate using a Ramsey echo experiment, Fig. \ref{supp_flux_noise}(a). The noise is applied only in the first half of the experiment to echo away any noise except the applied flux noise. Each echo sequence is repeated 50 times with different randomized flux-noise pulses.  Then the dephasing rate, $\gamma_\phi$, is fit to each curve, Fig. \ref{supp_flux_noise}(c).  Finally, a quadratic function is used to fit $\gamma_\phi$ as a function of flux-pulse amplitude.  

	\section{Fermi's Golden Rule and the Kofman-Kurizki Formula}
	\subsection{Background}
	
	Fermi's golden rule (FGR) tells us that the decay rate of a qubit into its bath, $\gamma_q$, is determined by the bath's spectral density, $\mathcal{B}(\omega)$, at the qubit frequency, $\gamma_q(\omega_q) = g_{SB}^2\mathcal{B}(\omega_q)$ \cite{clerk2010introduction}, where $g_{SB}$ is the interaction strength (formally the matrix element connecting the relevant qubit states) of the qubit-bath interaction.

	In the main text we extend the form of the Kofman-Kurizki formula to the case of dispersive readout, so here we review the general form of the equation.  
	The Kofman-Kurizki universal formula for the decay rate, $\Gamma$, of a qubit coupled to a bath is 
	\[ \Gamma = 2\pi \int_{-\infty}^{\infty} F(\omega) G(\omega) \,d\omega , \]
	where $G(\omega) = g_{SB}^2\mathcal{B}(\omega)/(2\pi)$ and $F(\omega)$ is the measurement-induced qubit-broadening function that depends on the type and parameters of the measurement \cite{kofman2000acceleration, kofman2001frequent, kofman2001zeno}. 
	In the case of continuous measurement the qubit-broadening function,
	\[ F(\omega) =  \frac{1}{\pi}\frac{\gamma_m}{\gamma_m^2 + \left(\omega -  \omega_q  \right)^2}, \]
	is a Lorentzian with a width given by the measurement rate, $\gamma_m$.
	As the measurement rate goes to zero, $\gamma_m \rightarrow 0$, the Lorentzian becomes a delta function at the qubit frequency, $\delta(\omega_q)$, so the decay rate becomes $\Gamma \rightarrow 2 \pi G(\omega_q) = g_{SB}^2\mathcal{B}(\omega_q)$, which we recognize as Fermi's golden rule \cite{clerk2010introduction}.
	
	\subsection{The Purcell effect and the Zeno effect}
	
	The deep connections between the Zeno effect and the Purcell effect can see seen most easily by looking at the Purcell loss of a qubit that is resonant with a readout resonator, as shown in Fig.~1 of the main text. When the qubit's coupling to the resonator is small compared to the decay rate of the resonator, $g \ll \kappa$, then the Purcell loss on the qubit is $\gamma_P = 4 g^2 / \kappa$, which has the arguably counter-intuitive feature that increasing the decay rate of the resonator, $\kappa$, decreases the induced decay rate of the qubit, $\gamma_P$ \cite{girvin2014circuit}. Because any photon emitted from the cavity can in-principle be detected, $\kappa$ is a measurement rate, which suggests an interpretation of this in terms of the Zeno-effect.  
	
	A classic problem demonstrating the Zeno effect is a qubit that is simultaneously driven at Rabi rate $\Omega$ and continuously measured at rate $\gamma_M$.  When the measurement rate is much faster than the Rabi rate, $\Omega \ll \gamma_M$, the measurement prevents the qubit from evolving. So rather than Rabi-evolution between the ground and excited states, the qubit will undergo quantum jumps between them at a rate $\Omega^2 / \gamma_M$ \cite{schulman1998continuous}.  Therefore, we can understand the dependence of the the Purcell loss on the decay rate of the readout resonator in the resonant limit in terms of the Zeno effect as an excitation in the qubit being prevented from swapping into the resonator (at vacuum Rabi rate 2$g$) by the Zeno effect due to the continuous measurement of the resonator at rate $\kappa$.

	\subsection{Self-consistent quantum master equation derivation of the Kofman-Kurizki formula}

	In this section we outline the derivation of the Kofman-Kurizki formula using the Dissipation Incorporated Self-Consistent (DISCo) coarse graining approach. First introduced in Ref.~\cite{DISCO}, DISCo incorporates the impact of intrinsic system dissipation into the derivation of a Lindblad form master equation describing the evolution of the system due to interaction with a second, auxiliary bath. This is exactly the situation of the Kofman-Kurizki formula, where dephasing due to measurement modifies the relaxation of the qubit into its radiative environment. The full development of the DISCo formalism, and derivation of the Kofman-Kurizki formula are contained in Ref.~\cite{DISCO}, and for brevity, we only sketch the derivation here. Interested readers should refer to Ref.~\cite{DISCO} for further details.
	
	The ensemble average evolution of a qubit under continuous measurement can be described by the master equation
	\begin{align}
		\dot{\rho} = \mathcal{L}_{M}\rho = -i\left[\frac{\tilde{\omega}_q}{2}\hat{\sigma}_z,\rho\right] + \frac{\gamma_\phi}{2}\mathcal{D}\left[\hat{\sigma}_z\right]\rho,
	\end{align}
	where $\mathcal{D}[x]\bullet = x\bullet x^\dagger - \left\{x^\dagger x,\bullet\right\}/2$ is the usual dissipator. This description is completely general, with no consideration for the physical mechanism behind the measurement (or dephasing). For our particular case of dispersive readout, as in the main text we identify that $\tilde{\omega}_q$ includes the Stark-shift on the bare qubit frequency $\omega_q$ due to readout resonator photons, and the dephasing rate $\gamma_\phi$ is a function of photon number.
	
	We now consider the transversal interaction of the qubit with an auxiliary bath, which we identify as its radiative environment, via the Hamiltonian
	\begin{align}
		\hat{H}_{SB} = g_{SB}\hat{\sigma}_x \otimes \sum_j \mu_j \left(\hat{b}_j + \hat{b}_j^\dagger\right),
	\end{align}
	where $\hat{b}_j$ are the bath modes, $g_{SB}$ the average interaction strength, and $\mu_j$ are dimensionless real numbers. The standard derivation of the Lindblad master equation through the Born-Markov method would start by defining an interaction-frame system-bath Hamiltonian in the unitary frame defined by $\hat{H} = \tilde{\omega}_q\hat{\sigma}_z/2 + \hat{H}_B$, where $\hat{H}_B$ is the auxiliary bath self-Hamiltonian. To include the impact of the qubit dephasing, DISCo instead uses an interaction-frame that is obtained by transforming the system-bath Hamiltonian by an operator that is completely-positive and trace-preserving but not unitary. This transformation operator is defined by the generator $\mathcal{L} = \mathcal{L}_M + \mathcal{L}_B$, where $\mathcal{L}_B$ is the superoperator description of $\hat{H}_B$.
	
	Given this change of starting frame, the derivation of the DISCo master equation follows the standard Born-Markov rotating-wave-approximation methodology \cite{Lidar}, with a few additional terms dropped under the assumption that the intrinsic bath dynamics equilibrate faster than $\gamma_\phi$ \cite{DISCO}. The end result is that the DISCo master equation differs from the standard one in the form of the single-sided Fourier transform that defines the decay rate and the Lamb shift, but the structure of the resultant master equation is the same
	\begin{align}
		\dot{\rho} =  -i\left[\hat{H}_S,\rho\right] + \frac{\gamma_\phi}{2}\mathcal{D}\left[\hat{\sigma}_z\right]\rho + \Gamma(\tilde{\omega}_q)\mathcal{D}\left[\hat{\sigma}_-\right]\rho, \label{eqn:KKDISCO}
	\end{align}
	where $\hat{H}_S$ includes the system self-Hamiltonian and the Lamb shift.
	
	Focusing on the decay rate, for DISCo this is given by the convolution of the bath spectral density with a Lorentzian filter describing the impact of finite qubit dephasing
	\begin{align}
		\Gamma(\tilde{\omega}_q) = \frac{g_{SB}^2}{\pi}\int_{-\infty}^{\infty} \mathcal{B}(\omega) \frac{\gamma_\phi}{\gamma_\phi^2 + \left(\omega - \tilde{\omega}_q \right)^2} \,d\omega,
	\end{align}
	where $\mathcal{B}(\omega)$ is the bath spectral density. This is exactly the Kofman-Kurizki formula for continuous measurement. We recover the expression of the main text with the identifications $\tilde{\omega}_q = \omega_q + \nu_S(\bar{n}_r)$, $\gamma_\phi = \gamma_\phi(\bar{n}_r)$, and $\gamma_q(\omega) = g_{SB}^2\mathcal{B}(\omega)$. The expression for $\gamma_q(\omega)$ uses the assumption that without measurement the decay rate is simply given by the Fermi's golden rule expression at a given frequency $\omega$.
	
	It is important to note that while we have used this derivation of the Kofman-Kurizki formula to model lifetime reduction of a superconducting qubit during readout, we have not explicitly included the readout resonator in our derivation, only its effect (Stark shift and dephasing) on the qubit. This is the key distinction between our results and previous attempts to explain lifetime reduction that explored how driven-resonator modifications to the qubit-resonator interaction either change the Purcell decay rate of the qubit \cite{sete2014purcell, malekakhlagh2020lifetime, petrescu2020lifetime, hanai2021intrinsic, boissonneault2010improved, sete2015quantum, muller2020dissipative}, introduce new dissipative pathways \cite{boissonneault2008nonlinear, boissonneault2009dispersive, slichter2012measurement}, or activate previously off-resonant transitions \cite{sank2016measurement, khezri2022measurement, shillito2022dynamics, cohen2022reminiscence, verney2019structural, lescanne2019escape}. Here, rather than focusing on how the qubit-resonator interaction is modified, we explore how measurement modifies the qubit's decay into the rest of its radiative environment.
	
	We expect our results to be valid when it is safe to neglect the coherent interaction between the qubit and readout resonator, such that we can treat the readout resonator solely as a source of dephasing and Stark shift. The latter assumption is valid outside of the number splitting regime, i.e.~$\chi \ll \kappa$, and for our device $\chi/2\pi$ = 0.98~MHz and $\kappa/2\pi$ = 5.24~MHz. As $\chi$ becomes a more appreciable fraction of $\kappa$, it may be possible to extend our results using more complete descriptions of the qubit absorption spectrum \cite{gambetta2006qubit}. When Purcell decay becomes significant, we expect a combination of our results and those from previous work to be applicable.  Also, we have not accounted for the time-dynamics of the induced qubit dissipation during ring-up and ring-down of the readout resonator, as well as other transient effects that could occur. Understanding these intermediate regimes and dynamical effects is beyond the scope of this work and is a topic for future study.
	
	\subsection{Kofman-Kurizki formula in the single-TLS limit}
	
	Ref.~\cite{DISCO} provides a technique to derive a self-consistent quantum master equations in the limit where the ``bath'' of the system consists of a strongly-coupled defect that dissipates into its own environment. This is the situation we have studied with the artificial-TLS experiments experiments in the main text, and Eq. 2 of the main text was taken from Ref.~\cite{DISCO}.  We dropped small terms that depended on the sum of $\omega_D$ and $\omega_q$ rather than the difference, including a small heating term.  This form also differs from the form in Ref.~\cite{DISCO} in that we have substituted $\gamma_\phi = 2 \lambda_q$, due to a difference in the notation for the dephasing operator.
	While Ref.~\cite{DISCO} studies the more general situation of a system interacting with a unipartite bath, for the particular setup of this work most of Eq. 2 of the main text can be obtained directly from the Kofman-Kurizki formula.
	
	We model the defect transmon as a harmonic oscillator, i.e.~we ignore its anharmonicity, which for such a strongly damped transmon is sufficiently accurate for our purposes. For a transmon frequency $\omega_D$ and decay rate (into its own bath) $\gamma_{1,D}$, then its two-point correlation function is $\mathcal{B}(\tau) = e^{-i\omega_D\tau - \gamma_{1,D}|\tau|/2}$, such that its spectral density, the Fourier transform of $\mathcal{B}(\tau)$, is given by
	\begin{align}
		\mathcal{B}(\omega) = \frac{\gamma_{1,D}}{\left(\gamma_{1,D}/2\right)^2 + (\omega - \omega_D)^2}.
	\end{align}
	Plugging this into Eq.~\eqref{eqn:KKDISCO} and integrating the product of Lorentzians over all of frequency space gives the expression found in Eq. 2 excluding the impact of the qubit's decay rate $\gamma_q$. To include $\gamma_q$ we must derive a new master equation following the DISCo technique \cite{DISCO}.
	

\begin{thebibliography}{61}%
		\makeatletter
		\providecommand \@ifxundefined [1]{%
			\@ifx{#1\undefined}
		}%
		\providecommand \@ifnum [1]{%
			\ifnum #1\expandafter \@firstoftwo
			\else \expandafter \@secondoftwo
			\fi
		}%
		\providecommand \@ifx [1]{%
			\ifx #1\expandafter \@firstoftwo
			\else \expandafter \@secondoftwo
			\fi
		}%
		\providecommand \natexlab [1]{#1}%
		\providecommand \enquote  [1]{``#1''}%
		\providecommand \bibnamefont  [1]{#1}%
		\providecommand \bibfnamefont [1]{#1}%
		\providecommand \citenamefont [1]{#1}%
		\providecommand \href@noop [0]{\@secondoftwo}%
		\providecommand \href [0]{\begingroup \@sanitize@url \@href}%
		\providecommand \@href[1]{\@@startlink{#1}\@@href}%
		\providecommand \@@href[1]{\endgroup#1\@@endlink}%
		\providecommand \@sanitize@url [0]{\catcode `\\12\catcode `\$12\catcode
			`\&12\catcode `\#12\catcode `\^12\catcode `\_12\catcode `\%12\relax}%
		\providecommand \@@startlink[1]{}%
		\providecommand \@@endlink[0]{}%
		\providecommand \url  [0]{\begingroup\@sanitize@url \@url }%
		\providecommand \@url [1]{\endgroup\@href {#1}{\urlprefix }}%
		\providecommand \urlprefix  [0]{URL }%
		\providecommand \Eprint [0]{\href }%
		\providecommand \doibase [0]{https://doi.org/}%
		\providecommand \selectlanguage [0]{\@gobble}%
		\providecommand \bibinfo  [0]{\@secondoftwo}%
		\providecommand \bibfield  [0]{\@secondoftwo}%
		\providecommand \translation [1]{[#1]}%
		\providecommand \BibitemOpen [0]{}%
		\providecommand \bibitemStop [0]{}%
		\providecommand \bibitemNoStop [0]{.\EOS\space}%
		\providecommand \EOS [0]{\spacefactor3000\relax}%
		\providecommand \BibitemShut  [1]{\csname bibitem#1\endcsname}%
		\let\auto@bib@innerbib\@empty
		\bibitem [{\citenamefont {Blais}\ \emph {et~al.}(2004)\citenamefont {Blais},
			\citenamefont {Huang}, \citenamefont {Wallraff}, \citenamefont {Girvin},\
			and\ \citenamefont {Schoelkopf}}]{blais2004cavity}%
		\BibitemOpen
		\bibfield  {author} {\bibinfo {author} {\bibfnamefont {A.}~\bibnamefont
				{Blais}}, \bibinfo {author} {\bibfnamefont {R.-S.}\ \bibnamefont {Huang}},
			\bibinfo {author} {\bibfnamefont {A.}~\bibnamefont {Wallraff}}, \bibinfo
			{author} {\bibfnamefont {S.~M.}\ \bibnamefont {Girvin}},\ and\ \bibinfo
			{author} {\bibfnamefont {R.~J.}\ \bibnamefont {Schoelkopf}},\ }\href@noop {}
		{\bibfield  {journal} {\bibinfo  {journal} {Phys. Rev. A}\ }\textbf {\bibinfo
				{volume} {69}},\ \bibinfo {pages} {062320} (\bibinfo {year}
			{2004})}\BibitemShut {NoStop}%
		\bibitem [{\citenamefont {Blais}\ \emph {et~al.}(2021)\citenamefont {Blais},
			\citenamefont {Grimsmo}, \citenamefont {Girvin},\ and\ \citenamefont
			{Wallraff}}]{blais2021circuit}%
		\BibitemOpen
		\bibfield  {author} {\bibinfo {author} {\bibfnamefont {A.}~\bibnamefont
				{Blais}}, \bibinfo {author} {\bibfnamefont {A.~L.}\ \bibnamefont {Grimsmo}},
			\bibinfo {author} {\bibfnamefont {S.~M.}\ \bibnamefont {Girvin}},\ and\
			\bibinfo {author} {\bibfnamefont {A.}~\bibnamefont {Wallraff}},\ }\href@noop
		{} {\bibfield  {journal} {\bibinfo  {journal} {Rev. Mod. Phys.}\ }\textbf
			{\bibinfo {volume} {93}},\ \bibinfo {pages} {025005} (\bibinfo {year}
			{2021})}\BibitemShut {NoStop}%
		\bibitem [{\citenamefont {Wallraff}\ \emph {et~al.}(2005)\citenamefont
			{Wallraff}, \citenamefont {Schuster}, \citenamefont {Blais}, \citenamefont
			{Frunzio}, \citenamefont {Majer}, \citenamefont {Devoret}, \citenamefont
			{Girvin},\ and\ \citenamefont {Schoelkopf}}]{wallraff2005approaching}%
		\BibitemOpen
		\bibfield  {author} {\bibinfo {author} {\bibfnamefont {A.}~\bibnamefont
				{Wallraff}}, \bibinfo {author} {\bibfnamefont {D.~I.}\ \bibnamefont
				{Schuster}}, \bibinfo {author} {\bibfnamefont {A.}~\bibnamefont {Blais}},
			\bibinfo {author} {\bibfnamefont {L.}~\bibnamefont {Frunzio}}, \bibinfo
			{author} {\bibfnamefont {J.}~\bibnamefont {Majer}}, \bibinfo {author}
			{\bibfnamefont {M.~H.}\ \bibnamefont {Devoret}}, \bibinfo {author}
			{\bibfnamefont {S.~M.}\ \bibnamefont {Girvin}},\ and\ \bibinfo {author}
			{\bibfnamefont {R.~J.}\ \bibnamefont {Schoelkopf}},\ }\href@noop {}
		{\bibfield  {journal} {\bibinfo  {journal} {Phys. Rev. Lett.}\ }\textbf
			{\bibinfo {volume} {95}},\ \bibinfo {pages} {060501} (\bibinfo {year}
			{2005})}\BibitemShut {NoStop}%
		\bibitem [{\citenamefont {Bravyi}\ \emph {et~al.}(2021)\citenamefont {Bravyi},
			\citenamefont {Sheldon}, \citenamefont {Kandala}, \citenamefont {Mckay},\
			and\ \citenamefont {Gambetta}}]{bravyi2021mitigating}%
		\BibitemOpen
		\bibfield  {author} {\bibinfo {author} {\bibfnamefont {S.}~\bibnamefont
				{Bravyi}}, \bibinfo {author} {\bibfnamefont {S.}~\bibnamefont {Sheldon}},
			\bibinfo {author} {\bibfnamefont {A.}~\bibnamefont {Kandala}}, \bibinfo
			{author} {\bibfnamefont {D.~C.}\ \bibnamefont {Mckay}},\ and\ \bibinfo
			{author} {\bibfnamefont {J.~M.}\ \bibnamefont {Gambetta}},\ }\href@noop {}
		{\bibfield  {journal} {\bibinfo  {journal} {Phys. Rev. A}\ }\textbf {\bibinfo
				{volume} {103}},\ \bibinfo {pages} {042605} (\bibinfo {year}
			{2021})}\BibitemShut {NoStop}%
		\bibitem [{\citenamefont {Van Den~Berg}\ \emph {et~al.}(2022)\citenamefont {Van
				Den~Berg}, \citenamefont {Minev},\ and\ \citenamefont
			{Temme}}]{vandenberg2022model}%
		\BibitemOpen
		\bibfield  {author} {\bibinfo {author} {\bibfnamefont {E.}~\bibnamefont {Van
					Den~Berg}}, \bibinfo {author} {\bibfnamefont {Z.~K.}\ \bibnamefont {Minev}},\
			and\ \bibinfo {author} {\bibfnamefont {K.}~\bibnamefont {Temme}},\
		}\href@noop {} {\bibfield  {journal} {\bibinfo  {journal} {Phys. Rev. A}\
			}\textbf {\bibinfo {volume} {105}},\ \bibinfo {pages} {032620} (\bibinfo
			{year} {2022})}\BibitemShut {NoStop}%
		\bibitem [{\citenamefont {Fowler}\ \emph {et~al.}(2012)\citenamefont {Fowler},
			\citenamefont {Mariantoni}, \citenamefont {Martinis},\ and\ \citenamefont
			{Cleland}}]{fowler2012surface}%
		\BibitemOpen
		\bibfield  {author} {\bibinfo {author} {\bibfnamefont {A.~G.}\ \bibnamefont
				{Fowler}}, \bibinfo {author} {\bibfnamefont {M.}~\bibnamefont {Mariantoni}},
			\bibinfo {author} {\bibfnamefont {J.~M.}\ \bibnamefont {Martinis}},\ and\
			\bibinfo {author} {\bibfnamefont {A.~N.}\ \bibnamefont {Cleland}},\
		}\href@noop {} {\bibfield  {journal} {\bibinfo  {journal} {Phys. Rev. A}\
			}\textbf {\bibinfo {volume} {86}},\ \bibinfo {pages} {032324} (\bibinfo
			{year} {2012})}\BibitemShut {NoStop}%
		\bibitem [{\citenamefont {Leymann}\ and\ \citenamefont
			{Barzen}(2020)}]{leymann2020bitter}%
		\BibitemOpen
		\bibfield  {author} {\bibinfo {author} {\bibfnamefont {F.}~\bibnamefont
				{Leymann}}\ and\ \bibinfo {author} {\bibfnamefont {J.}~\bibnamefont
				{Barzen}},\ }\href@noop {} {\bibfield  {journal} {\bibinfo  {journal}
				{Quantum Sci. Technol.}\ }\textbf {\bibinfo {volume} {5}},\ \bibinfo {pages}
			{044007} (\bibinfo {year} {2020})}\BibitemShut {NoStop}%
		\bibitem [{\citenamefont {C{\'o}rcoles}\ \emph {et~al.}(2021)\citenamefont
			{C{\'o}rcoles}, \citenamefont {Takita}, \citenamefont {Inoue}, \citenamefont
			{Lekuch}, \citenamefont {Minev}, \citenamefont {Chow},\ and\ \citenamefont
			{Gambetta}}]{corcoles2021exploiting}%
		\BibitemOpen
		\bibfield  {author} {\bibinfo {author} {\bibfnamefont {A.~D.}\ \bibnamefont
				{C{\'o}rcoles}}, \bibinfo {author} {\bibfnamefont {M.}~\bibnamefont
				{Takita}}, \bibinfo {author} {\bibfnamefont {K.}~\bibnamefont {Inoue}},
			\bibinfo {author} {\bibfnamefont {S.}~\bibnamefont {Lekuch}}, \bibinfo
			{author} {\bibfnamefont {Z.~K.}\ \bibnamefont {Minev}}, \bibinfo {author}
			{\bibfnamefont {J.~M.}\ \bibnamefont {Chow}},\ and\ \bibinfo {author}
			{\bibfnamefont {J.~M.}\ \bibnamefont {Gambetta}},\ }\href@noop {} {\bibfield
			{journal} {\bibinfo  {journal} {Phys. Rev. Lett.}\ }\textbf {\bibinfo
				{volume} {127}},\ \bibinfo {pages} {100501} (\bibinfo {year}
			{2021})}\BibitemShut {NoStop}%
		\bibitem [{\citenamefont {Sank}\ \emph {et~al.}(2016)\citenamefont {Sank},
			\citenamefont {Chen}, \citenamefont {Khezri}, \citenamefont {Kelly},
			\citenamefont {Barends}, \citenamefont {Campbell}, \citenamefont {Chen},
			\citenamefont {Chiaro}, \citenamefont {Dunsworth}, \citenamefont {Fowler}
			\emph {et~al.}}]{sank2016measurement}%
		\BibitemOpen
		\bibfield  {author} {\bibinfo {author} {\bibfnamefont {D.}~\bibnamefont
				{Sank}}, \bibinfo {author} {\bibfnamefont {Z.}~\bibnamefont {Chen}}, \bibinfo
			{author} {\bibfnamefont {M.}~\bibnamefont {Khezri}}, \bibinfo {author}
			{\bibfnamefont {J.}~\bibnamefont {Kelly}}, \bibinfo {author} {\bibfnamefont
				{R.}~\bibnamefont {Barends}}, \bibinfo {author} {\bibfnamefont
				{B.}~\bibnamefont {Campbell}}, \bibinfo {author} {\bibfnamefont
				{Y.}~\bibnamefont {Chen}}, \bibinfo {author} {\bibfnamefont {B.}~\bibnamefont
				{Chiaro}}, \bibinfo {author} {\bibfnamefont {A.}~\bibnamefont {Dunsworth}},
			\bibinfo {author} {\bibfnamefont {A.}~\bibnamefont {Fowler}}, \emph
			{et~al.},\ }\href@noop {} {\bibfield  {journal} {\bibinfo  {journal} {Phys.
					Rev. Lett.}\ }\textbf {\bibinfo {volume} {117}},\ \bibinfo {pages} {190503}
			(\bibinfo {year} {2016})}\BibitemShut {NoStop}%
		\bibitem [{\citenamefont {Khezri}\ \emph {et~al.}(2022)\citenamefont {Khezri},
			\citenamefont {Opremcak}, \citenamefont {Chen}, \citenamefont {Bengtsson},
			\citenamefont {White}, \citenamefont {Naaman}, \citenamefont {Acharya},
			\citenamefont {Anderson}, \citenamefont {Ansmann}, \citenamefont {Arute}
			\emph {et~al.}}]{khezri2022measurement}%
		\BibitemOpen
		\bibfield  {author} {\bibinfo {author} {\bibfnamefont {M.}~\bibnamefont
				{Khezri}}, \bibinfo {author} {\bibfnamefont {A.}~\bibnamefont {Opremcak}},
			\bibinfo {author} {\bibfnamefont {Z.}~\bibnamefont {Chen}}, \bibinfo {author}
			{\bibfnamefont {A.}~\bibnamefont {Bengtsson}}, \bibinfo {author}
			{\bibfnamefont {T.}~\bibnamefont {White}}, \bibinfo {author} {\bibfnamefont
				{O.}~\bibnamefont {Naaman}}, \bibinfo {author} {\bibfnamefont
				{R.}~\bibnamefont {Acharya}}, \bibinfo {author} {\bibfnamefont
				{K.}~\bibnamefont {Anderson}}, \bibinfo {author} {\bibfnamefont
				{M.}~\bibnamefont {Ansmann}}, \bibinfo {author} {\bibfnamefont
				{F.}~\bibnamefont {Arute}}, \emph {et~al.},\ }\href@noop {} {\bibfield
			{journal} {\bibinfo  {journal} {arXiv:2212.05097}\ } (\bibinfo {year}
			{2022})}\BibitemShut {NoStop}%
		\bibitem [{\citenamefont {Shillito}\ \emph {et~al.}(2022)\citenamefont
			{Shillito}, \citenamefont {Petrescu}, \citenamefont {Cohen}, \citenamefont
			{Beall}, \citenamefont {Hauru}, \citenamefont {Ganahl}, \citenamefont
			{Lewis}, \citenamefont {Vidal},\ and\ \citenamefont
			{Blais}}]{shillito2022dynamics}%
		\BibitemOpen
		\bibfield  {author} {\bibinfo {author} {\bibfnamefont {R.}~\bibnamefont
				{Shillito}}, \bibinfo {author} {\bibfnamefont {A.}~\bibnamefont {Petrescu}},
			\bibinfo {author} {\bibfnamefont {J.}~\bibnamefont {Cohen}}, \bibinfo
			{author} {\bibfnamefont {J.}~\bibnamefont {Beall}}, \bibinfo {author}
			{\bibfnamefont {M.}~\bibnamefont {Hauru}}, \bibinfo {author} {\bibfnamefont
				{M.}~\bibnamefont {Ganahl}}, \bibinfo {author} {\bibfnamefont {A.~G.}\
				\bibnamefont {Lewis}}, \bibinfo {author} {\bibfnamefont {G.}~\bibnamefont
				{Vidal}},\ and\ \bibinfo {author} {\bibfnamefont {A.}~\bibnamefont {Blais}},\
		}\href@noop {} {\bibfield  {journal} {\bibinfo  {journal} {Phys. Rev. Appl.}\
			}\textbf {\bibinfo {volume} {18}},\ \bibinfo {pages} {034031} (\bibinfo
			{year} {2022})}\BibitemShut {NoStop}%
		\bibitem [{\citenamefont {Cohen}\ \emph {et~al.}(2022)\citenamefont {Cohen},
			\citenamefont {Petrescu}, \citenamefont {Shillito},\ and\ \citenamefont
			{Blais}}]{cohen2022reminiscence}%
		\BibitemOpen
		\bibfield  {author} {\bibinfo {author} {\bibfnamefont {J.}~\bibnamefont
				{Cohen}}, \bibinfo {author} {\bibfnamefont {A.}~\bibnamefont {Petrescu}},
			\bibinfo {author} {\bibfnamefont {R.}~\bibnamefont {Shillito}},\ and\
			\bibinfo {author} {\bibfnamefont {A.}~\bibnamefont {Blais}},\ }\href@noop {}
		{\bibfield  {journal} {\bibinfo  {journal} {arXiv:2207.09361}\ } (\bibinfo
			{year} {2022})}\BibitemShut {NoStop}%
		\bibitem [{\citenamefont {Verney}\ \emph {et~al.}(2019)\citenamefont {Verney},
			\citenamefont {Lescanne}, \citenamefont {Devoret}, \citenamefont {Leghtas},\
			and\ \citenamefont {Mirrahimi}}]{verney2019structural}%
		\BibitemOpen
		\bibfield  {author} {\bibinfo {author} {\bibfnamefont {L.}~\bibnamefont
				{Verney}}, \bibinfo {author} {\bibfnamefont {R.}~\bibnamefont {Lescanne}},
			\bibinfo {author} {\bibfnamefont {M.~H.}\ \bibnamefont {Devoret}}, \bibinfo
			{author} {\bibfnamefont {Z.}~\bibnamefont {Leghtas}},\ and\ \bibinfo {author}
			{\bibfnamefont {M.}~\bibnamefont {Mirrahimi}},\ }\href@noop {} {\bibfield
			{journal} {\bibinfo  {journal} {Phys. Rev. Appl.}\ }\textbf {\bibinfo
				{volume} {11}},\ \bibinfo {pages} {024003} (\bibinfo {year}
			{2019})}\BibitemShut {NoStop}%
		\bibitem [{\citenamefont {Lescanne}\ \emph {et~al.}(2019)\citenamefont
			{Lescanne}, \citenamefont {Verney}, \citenamefont {Ficheux}, \citenamefont
			{Devoret}, \citenamefont {Huard}, \citenamefont {Mirrahimi},\ and\
			\citenamefont {Leghtas}}]{lescanne2019escape}%
		\BibitemOpen
		\bibfield  {author} {\bibinfo {author} {\bibfnamefont {R.}~\bibnamefont
				{Lescanne}}, \bibinfo {author} {\bibfnamefont {L.}~\bibnamefont {Verney}},
			\bibinfo {author} {\bibfnamefont {Q.}~\bibnamefont {Ficheux}}, \bibinfo
			{author} {\bibfnamefont {M.~H.}\ \bibnamefont {Devoret}}, \bibinfo {author}
			{\bibfnamefont {B.}~\bibnamefont {Huard}}, \bibinfo {author} {\bibfnamefont
				{M.}~\bibnamefont {Mirrahimi}},\ and\ \bibinfo {author} {\bibfnamefont
				{Z.}~\bibnamefont {Leghtas}},\ }\href@noop {} {\bibfield  {journal} {\bibinfo
				{journal} {Phys. Rev. Appl.}\ }\textbf {\bibinfo {volume} {11}},\ \bibinfo
			{pages} {014030} (\bibinfo {year} {2019})}\BibitemShut {NoStop}%
		\bibitem [{\citenamefont {Malekakhlagh}\ \emph {et~al.}(2022)\citenamefont
			{Malekakhlagh}, \citenamefont {Shanks},\ and\ \citenamefont
			{Paik}}]{malekakhlagh2022optimization}%
		\BibitemOpen
		\bibfield  {author} {\bibinfo {author} {\bibfnamefont {M.}~\bibnamefont
				{Malekakhlagh}}, \bibinfo {author} {\bibfnamefont {W.}~\bibnamefont
				{Shanks}},\ and\ \bibinfo {author} {\bibfnamefont {H.}~\bibnamefont {Paik}},\
		}\href@noop {} {\bibfield  {journal} {\bibinfo  {journal} {Phys. Rev. A}\
			}\textbf {\bibinfo {volume} {105}},\ \bibinfo {pages} {022607} (\bibinfo
			{year} {2022})}\BibitemShut {NoStop}%
		\bibitem [{\citenamefont {Harrington}\ \emph {et~al.}(2017)\citenamefont
			{Harrington}, \citenamefont {Monroe},\ and\ \citenamefont
			{Murch}}]{harrington2017quantum}%
		\BibitemOpen
		\bibfield  {author} {\bibinfo {author} {\bibfnamefont {P.}~\bibnamefont
				{Harrington}}, \bibinfo {author} {\bibfnamefont {J.}~\bibnamefont {Monroe}},\
			and\ \bibinfo {author} {\bibfnamefont {K.}~\bibnamefont {Murch}},\
		}\href@noop {} {\bibfield  {journal} {\bibinfo  {journal} {Phys. Rev. Lett.}\
			}\textbf {\bibinfo {volume} {118}},\ \bibinfo {pages} {240401} (\bibinfo
			{year} {2017})}\BibitemShut {NoStop}%
		\bibitem [{\citenamefont {Johnson}\ \emph {et~al.}(2012)\citenamefont
			{Johnson}, \citenamefont {Macklin}, \citenamefont {Slichter}, \citenamefont
			{Vijay}, \citenamefont {Weingarten}, \citenamefont {Clarke},\ and\
			\citenamefont {Siddiqi}}]{johnson2012heralded}%
		\BibitemOpen
		\bibfield  {author} {\bibinfo {author} {\bibfnamefont {J.}~\bibnamefont
				{Johnson}}, \bibinfo {author} {\bibfnamefont {C.}~\bibnamefont {Macklin}},
			\bibinfo {author} {\bibfnamefont {D.}~\bibnamefont {Slichter}}, \bibinfo
			{author} {\bibfnamefont {R.}~\bibnamefont {Vijay}}, \bibinfo {author}
			{\bibfnamefont {E.}~\bibnamefont {Weingarten}}, \bibinfo {author}
			{\bibfnamefont {J.}~\bibnamefont {Clarke}},\ and\ \bibinfo {author}
			{\bibfnamefont {I.}~\bibnamefont {Siddiqi}},\ }\href@noop {} {\bibfield
			{journal} {\bibinfo  {journal} {Phys. Rev. Lett.}\ }\textbf {\bibinfo
				{volume} {109}},\ \bibinfo {pages} {050506} (\bibinfo {year}
			{2012})}\BibitemShut {NoStop}%
		\bibitem [{\citenamefont {Minev}\ \emph {et~al.}(2019)\citenamefont {Minev},
			\citenamefont {Mundhada}, \citenamefont {Shankar}, \citenamefont {Reinhold},
			\citenamefont {Guti{\'e}rrez-J{\'a}uregui}, \citenamefont {Schoelkopf},
			\citenamefont {Mirrahimi}, \citenamefont {Carmichael},\ and\ \citenamefont
			{Devoret}}]{minev2019catch}%
		\BibitemOpen
		\bibfield  {author} {\bibinfo {author} {\bibfnamefont {Z.~K.}\ \bibnamefont
				{Minev}}, \bibinfo {author} {\bibfnamefont {S.~O.}\ \bibnamefont {Mundhada}},
			\bibinfo {author} {\bibfnamefont {S.}~\bibnamefont {Shankar}}, \bibinfo
			{author} {\bibfnamefont {P.}~\bibnamefont {Reinhold}}, \bibinfo {author}
			{\bibfnamefont {R.}~\bibnamefont {Guti{\'e}rrez-J{\'a}uregui}}, \bibinfo
			{author} {\bibfnamefont {R.~J.}\ \bibnamefont {Schoelkopf}}, \bibinfo
			{author} {\bibfnamefont {M.}~\bibnamefont {Mirrahimi}}, \bibinfo {author}
			{\bibfnamefont {H.~J.}\ \bibnamefont {Carmichael}},\ and\ \bibinfo {author}
			{\bibfnamefont {M.~H.}\ \bibnamefont {Devoret}},\ }\href@noop {} {\bibfield
			{journal} {\bibinfo  {journal} {Nature}\ }\textbf {\bibinfo {volume} {570}},\
			\bibinfo {pages} {200} (\bibinfo {year} {2019})}\BibitemShut {NoStop}%
		\bibitem [{\citenamefont {Gusenkova}\ \emph {et~al.}(2021)\citenamefont
			{Gusenkova}, \citenamefont {Spiecker}, \citenamefont {Gebauer}, \citenamefont
			{Willsch}, \citenamefont {Willsch}, \citenamefont {Valenti}, \citenamefont
			{Karcher}, \citenamefont {Gr{\"u}nhaupt}, \citenamefont {Takmakov},
			\citenamefont {Winkel} \emph {et~al.}}]{gusenkova2021quantum}%
		\BibitemOpen
		\bibfield  {author} {\bibinfo {author} {\bibfnamefont {D.}~\bibnamefont
				{Gusenkova}}, \bibinfo {author} {\bibfnamefont {M.}~\bibnamefont {Spiecker}},
			\bibinfo {author} {\bibfnamefont {R.}~\bibnamefont {Gebauer}}, \bibinfo
			{author} {\bibfnamefont {M.}~\bibnamefont {Willsch}}, \bibinfo {author}
			{\bibfnamefont {D.}~\bibnamefont {Willsch}}, \bibinfo {author} {\bibfnamefont
				{F.}~\bibnamefont {Valenti}}, \bibinfo {author} {\bibfnamefont
				{N.}~\bibnamefont {Karcher}}, \bibinfo {author} {\bibfnamefont
				{L.}~\bibnamefont {Gr{\"u}nhaupt}}, \bibinfo {author} {\bibfnamefont
				{I.}~\bibnamefont {Takmakov}}, \bibinfo {author} {\bibfnamefont
				{P.}~\bibnamefont {Winkel}}, \emph {et~al.},\ }\href@noop {} {\bibfield
			{journal} {\bibinfo  {journal} {Phys. Rev. Appl.}\ }\textbf {\bibinfo
				{volume} {15}},\ \bibinfo {pages} {064030} (\bibinfo {year}
			{2021})}\BibitemShut {NoStop}%
		\bibitem [{\citenamefont {Sivak}\ \emph {et~al.}(2023)\citenamefont {Sivak},
			\citenamefont {Eickbusch}, \citenamefont {Royer}, \citenamefont {Singh},
			\citenamefont {Tsioutsios}, \citenamefont {Ganjam}, \citenamefont {Miano},
			\citenamefont {Brock}, \citenamefont {Ding}, \citenamefont {Frunzio} \emph
			{et~al.}}]{sivak2022real}%
		\BibitemOpen
		\bibfield  {author} {\bibinfo {author} {\bibfnamefont {V.}~\bibnamefont
				{Sivak}}, \bibinfo {author} {\bibfnamefont {A.}~\bibnamefont {Eickbusch}},
			\bibinfo {author} {\bibfnamefont {B.}~\bibnamefont {Royer}}, \bibinfo
			{author} {\bibfnamefont {S.}~\bibnamefont {Singh}}, \bibinfo {author}
			{\bibfnamefont {I.}~\bibnamefont {Tsioutsios}}, \bibinfo {author}
			{\bibfnamefont {S.}~\bibnamefont {Ganjam}}, \bibinfo {author} {\bibfnamefont
				{A.}~\bibnamefont {Miano}}, \bibinfo {author} {\bibfnamefont
				{B.}~\bibnamefont {Brock}}, \bibinfo {author} {\bibfnamefont
				{A.}~\bibnamefont {Ding}}, \bibinfo {author} {\bibfnamefont {L.}~\bibnamefont
				{Frunzio}}, \emph {et~al.},\ }\href@noop {} {\bibfield  {journal} {\bibinfo
				{journal} {Nature}\ } (\bibinfo {year} {2023})}\BibitemShut {NoStop}%
		\bibitem [{\citenamefont {Mallet}\ \emph {et~al.}(2009)\citenamefont {Mallet},
			\citenamefont {Ong}, \citenamefont {Palacios-Laloy}, \citenamefont {Nguyen},
			\citenamefont {Bertet}, \citenamefont {Vion},\ and\ \citenamefont
			{Esteve}}]{mallet2009single}%
		\BibitemOpen
		\bibfield  {author} {\bibinfo {author} {\bibfnamefont {F.}~\bibnamefont
				{Mallet}}, \bibinfo {author} {\bibfnamefont {F.~R.}\ \bibnamefont {Ong}},
			\bibinfo {author} {\bibfnamefont {A.}~\bibnamefont {Palacios-Laloy}},
			\bibinfo {author} {\bibfnamefont {F.}~\bibnamefont {Nguyen}}, \bibinfo
			{author} {\bibfnamefont {P.}~\bibnamefont {Bertet}}, \bibinfo {author}
			{\bibfnamefont {D.}~\bibnamefont {Vion}},\ and\ \bibinfo {author}
			{\bibfnamefont {D.}~\bibnamefont {Esteve}},\ }\href@noop {} {\bibfield
			{journal} {\bibinfo  {journal} {Nat. Phys.}\ }\textbf {\bibinfo {volume}
				{5}},\ \bibinfo {pages} {791} (\bibinfo {year} {2009})}\BibitemShut {NoStop}%
		\bibitem [{\citenamefont {Picot}\ \emph {et~al.}(2008)\citenamefont {Picot},
			\citenamefont {Lupa{\c{s}}cu}, \citenamefont {Saito}, \citenamefont
			{Harmans},\ and\ \citenamefont {Mooij}}]{picot2008role}%
		\BibitemOpen
		\bibfield  {author} {\bibinfo {author} {\bibfnamefont {T.}~\bibnamefont
				{Picot}}, \bibinfo {author} {\bibfnamefont {A.}~\bibnamefont
				{Lupa{\c{s}}cu}}, \bibinfo {author} {\bibfnamefont {S.}~\bibnamefont
				{Saito}}, \bibinfo {author} {\bibfnamefont {C.}~\bibnamefont {Harmans}},\
			and\ \bibinfo {author} {\bibfnamefont {J.}~\bibnamefont {Mooij}},\
		}\href@noop {} {\bibfield  {journal} {\bibinfo  {journal} {Phys. Rev. B}\
			}\textbf {\bibinfo {volume} {78}},\ \bibinfo {pages} {132508} (\bibinfo
			{year} {2008})}\BibitemShut {NoStop}%
		\bibitem [{\citenamefont {Serban}\ \emph {et~al.}(2010)\citenamefont {Serban},
			\citenamefont {Dykman},\ and\ \citenamefont
			{Wilhelm}}]{serban2010relaxation}%
		\BibitemOpen
		\bibfield  {author} {\bibinfo {author} {\bibfnamefont {I.}~\bibnamefont
				{Serban}}, \bibinfo {author} {\bibfnamefont {M.}~\bibnamefont {Dykman}},\
			and\ \bibinfo {author} {\bibfnamefont {F.}~\bibnamefont {Wilhelm}},\
		}\href@noop {} {\bibfield  {journal} {\bibinfo  {journal} {Phys. Rev. A}\
			}\textbf {\bibinfo {volume} {81}},\ \bibinfo {pages} {022305} (\bibinfo
			{year} {2010})}\BibitemShut {NoStop}%
		\bibitem [{\citenamefont {Boissonneault}\ \emph {et~al.}(2008)\citenamefont
			{Boissonneault}, \citenamefont {Gambetta},\ and\ \citenamefont
			{Blais}}]{boissonneault2008nonlinear}%
		\BibitemOpen
		\bibfield  {author} {\bibinfo {author} {\bibfnamefont {M.}~\bibnamefont
				{Boissonneault}}, \bibinfo {author} {\bibfnamefont {J.~M.}\ \bibnamefont
				{Gambetta}},\ and\ \bibinfo {author} {\bibfnamefont {A.}~\bibnamefont
				{Blais}},\ }\href@noop {} {\bibfield  {journal} {\bibinfo  {journal} {Phys.
					Rev. A}\ }\textbf {\bibinfo {volume} {77}},\ \bibinfo {pages} {060305}
			(\bibinfo {year} {2008})}\BibitemShut {NoStop}%
		\bibitem [{\citenamefont {Boissonneault}\ \emph {et~al.}(2009)\citenamefont
			{Boissonneault}, \citenamefont {Gambetta},\ and\ \citenamefont
			{Blais}}]{boissonneault2009dispersive}%
		\BibitemOpen
		\bibfield  {author} {\bibinfo {author} {\bibfnamefont {M.}~\bibnamefont
				{Boissonneault}}, \bibinfo {author} {\bibfnamefont {J.~M.}\ \bibnamefont
				{Gambetta}},\ and\ \bibinfo {author} {\bibfnamefont {A.}~\bibnamefont
				{Blais}},\ }\href@noop {} {\bibfield  {journal} {\bibinfo  {journal} {Phys.
					Rev. A}\ }\textbf {\bibinfo {volume} {79}},\ \bibinfo {pages} {013819}
			(\bibinfo {year} {2009})}\BibitemShut {NoStop}%
		\bibitem [{\citenamefont {Slichter}\ \emph {et~al.}(2012)\citenamefont
			{Slichter}, \citenamefont {Vijay}, \citenamefont {Weber}, \citenamefont
			{Boutin}, \citenamefont {Boissonneault}, \citenamefont {Gambetta},
			\citenamefont {Blais},\ and\ \citenamefont
			{Siddiqi}}]{slichter2012measurement}%
		\BibitemOpen
		\bibfield  {author} {\bibinfo {author} {\bibfnamefont {D.}~\bibnamefont
				{Slichter}}, \bibinfo {author} {\bibfnamefont {R.}~\bibnamefont {Vijay}},
			\bibinfo {author} {\bibfnamefont {S.}~\bibnamefont {Weber}}, \bibinfo
			{author} {\bibfnamefont {S.}~\bibnamefont {Boutin}}, \bibinfo {author}
			{\bibfnamefont {M.}~\bibnamefont {Boissonneault}}, \bibinfo {author}
			{\bibfnamefont {J.~M.}\ \bibnamefont {Gambetta}}, \bibinfo {author}
			{\bibfnamefont {A.}~\bibnamefont {Blais}},\ and\ \bibinfo {author}
			{\bibfnamefont {I.}~\bibnamefont {Siddiqi}},\ }\href@noop {} {\bibfield
			{journal} {\bibinfo  {journal} {Phys. Rev. Lett.}\ }\textbf {\bibinfo
				{volume} {109}},\ \bibinfo {pages} {153601} (\bibinfo {year}
			{2012})}\BibitemShut {NoStop}%
		\bibitem [{\citenamefont {Sete}\ \emph {et~al.}(2014)\citenamefont {Sete},
			\citenamefont {Gambetta},\ and\ \citenamefont {Korotkov}}]{sete2014purcell}%
		\BibitemOpen
		\bibfield  {author} {\bibinfo {author} {\bibfnamefont {E.~A.}\ \bibnamefont
				{Sete}}, \bibinfo {author} {\bibfnamefont {J.~M.}\ \bibnamefont {Gambetta}},\
			and\ \bibinfo {author} {\bibfnamefont {A.~N.}\ \bibnamefont {Korotkov}},\
		}\href@noop {} {\bibfield  {journal} {\bibinfo  {journal} {Phys. Rev. B}\
			}\textbf {\bibinfo {volume} {89}},\ \bibinfo {pages} {104516} (\bibinfo
			{year} {2014})}\BibitemShut {NoStop}%
		\bibitem [{\citenamefont {Malekakhlagh}\ \emph {et~al.}(2020)\citenamefont
			{Malekakhlagh}, \citenamefont {Petrescu},\ and\ \citenamefont
			{T{\"u}reci}}]{malekakhlagh2020lifetime}%
		\BibitemOpen
		\bibfield  {author} {\bibinfo {author} {\bibfnamefont {M.}~\bibnamefont
				{Malekakhlagh}}, \bibinfo {author} {\bibfnamefont {A.}~\bibnamefont
				{Petrescu}},\ and\ \bibinfo {author} {\bibfnamefont {H.~E.}\ \bibnamefont
				{T{\"u}reci}},\ }\href@noop {} {\bibfield  {journal} {\bibinfo  {journal}
				{Phys. Rev. B}\ }\textbf {\bibinfo {volume} {101}},\ \bibinfo {pages}
			{134509} (\bibinfo {year} {2020})}\BibitemShut {NoStop}%
		\bibitem [{\citenamefont {Petrescu}\ \emph {et~al.}(2020)\citenamefont
			{Petrescu}, \citenamefont {Malekakhlagh},\ and\ \citenamefont
			{T{\"u}reci}}]{petrescu2020lifetime}%
		\BibitemOpen
		\bibfield  {author} {\bibinfo {author} {\bibfnamefont {A.}~\bibnamefont
				{Petrescu}}, \bibinfo {author} {\bibfnamefont {M.}~\bibnamefont
				{Malekakhlagh}},\ and\ \bibinfo {author} {\bibfnamefont {H.~E.}\ \bibnamefont
				{T{\"u}reci}},\ }\href@noop {} {\bibfield  {journal} {\bibinfo  {journal}
				{Phys. Rev. B}\ }\textbf {\bibinfo {volume} {101}},\ \bibinfo {pages}
			{134510} (\bibinfo {year} {2020})}\BibitemShut {NoStop}%
		\bibitem [{\citenamefont {Hanai}\ \emph {et~al.}(2021)\citenamefont {Hanai},
			\citenamefont {McDonald},\ and\ \citenamefont {Clerk}}]{hanai2021intrinsic}%
		\BibitemOpen
		\bibfield  {author} {\bibinfo {author} {\bibfnamefont {R.}~\bibnamefont
				{Hanai}}, \bibinfo {author} {\bibfnamefont {A.}~\bibnamefont {McDonald}},\
			and\ \bibinfo {author} {\bibfnamefont {A.}~\bibnamefont {Clerk}},\
		}\href@noop {} {\bibfield  {journal} {\bibinfo  {journal} {Phys. Rev. Res.}\
			}\textbf {\bibinfo {volume} {3}},\ \bibinfo {pages} {043228} (\bibinfo {year}
			{2021})}\BibitemShut {NoStop}%
		\bibitem [{\citenamefont {Boissonneault}\ \emph {et~al.}(2010)\citenamefont
			{Boissonneault}, \citenamefont {Gambetta},\ and\ \citenamefont
			{Blais}}]{boissonneault2010improved}%
		\BibitemOpen
		\bibfield  {author} {\bibinfo {author} {\bibfnamefont {M.}~\bibnamefont
				{Boissonneault}}, \bibinfo {author} {\bibfnamefont {J.}~\bibnamefont
				{Gambetta}},\ and\ \bibinfo {author} {\bibfnamefont {A.}~\bibnamefont
				{Blais}},\ }\href@noop {} {\bibfield  {journal} {\bibinfo  {journal} {Phys.
					Rev. Lett.}\ }\textbf {\bibinfo {volume} {105}},\ \bibinfo {pages} {100504}
			(\bibinfo {year} {2010})}\BibitemShut {NoStop}%
		\bibitem [{\citenamefont {Sete}\ \emph {et~al.}(2015)\citenamefont {Sete},
			\citenamefont {Martinis},\ and\ \citenamefont {Korotkov}}]{sete2015quantum}%
		\BibitemOpen
		\bibfield  {author} {\bibinfo {author} {\bibfnamefont {E.~A.}\ \bibnamefont
				{Sete}}, \bibinfo {author} {\bibfnamefont {J.~M.}\ \bibnamefont {Martinis}},\
			and\ \bibinfo {author} {\bibfnamefont {A.~N.}\ \bibnamefont {Korotkov}},\
		}\href@noop {} {\bibfield  {journal} {\bibinfo  {journal} {Phys. Rev. A}\
			}\textbf {\bibinfo {volume} {92}},\ \bibinfo {pages} {012325} (\bibinfo
			{year} {2015})}\BibitemShut {NoStop}%
		\bibitem [{\citenamefont {M{\"u}ller}(2020)}]{muller2020dissipative}%
		\BibitemOpen
		\bibfield  {author} {\bibinfo {author} {\bibfnamefont {C.}~\bibnamefont
				{M{\"u}ller}},\ }\href@noop {} {\bibfield  {journal} {\bibinfo  {journal}
				{Phys. Rev. Res.}\ }\textbf {\bibinfo {volume} {2}},\ \bibinfo {pages}
			{033046} (\bibinfo {year} {2020})}\BibitemShut {NoStop}%
		\bibitem [{\citenamefont {Reed}\ \emph {et~al.}(2010)\citenamefont {Reed},
			\citenamefont {Johnson}, \citenamefont {Houck}, \citenamefont {DiCarlo},
			\citenamefont {Chow}, \citenamefont {Schuster}, \citenamefont {Frunzio},\
			and\ \citenamefont {Schoelkopf}}]{reed2010fast}%
		\BibitemOpen
		\bibfield  {author} {\bibinfo {author} {\bibfnamefont {M.~D.}\ \bibnamefont
				{Reed}}, \bibinfo {author} {\bibfnamefont {B.~R.}\ \bibnamefont {Johnson}},
			\bibinfo {author} {\bibfnamefont {A.~A.}\ \bibnamefont {Houck}}, \bibinfo
			{author} {\bibfnamefont {L.}~\bibnamefont {DiCarlo}}, \bibinfo {author}
			{\bibfnamefont {J.~M.}\ \bibnamefont {Chow}}, \bibinfo {author}
			{\bibfnamefont {D.~I.}\ \bibnamefont {Schuster}}, \bibinfo {author}
			{\bibfnamefont {L.}~\bibnamefont {Frunzio}},\ and\ \bibinfo {author}
			{\bibfnamefont {R.~J.}\ \bibnamefont {Schoelkopf}},\ }\href@noop {}
		{\bibfield  {journal} {\bibinfo  {journal} {Appl. Phys. Lett.}\ }\textbf
			{\bibinfo {volume} {96}},\ \bibinfo {pages} {203110} (\bibinfo {year}
			{2010})}\BibitemShut {NoStop}%
		\bibitem [{\citenamefont {Jeffrey}\ \emph {et~al.}(2014)\citenamefont
			{Jeffrey}, \citenamefont {Sank}, \citenamefont {Mutus}, \citenamefont
			{White}, \citenamefont {Kelly}, \citenamefont {Barends}, \citenamefont
			{Chen}, \citenamefont {Chen}, \citenamefont {Chiaro}, \citenamefont
			{Dunsworth} \emph {et~al.}}]{jeffrey2014fast}%
		\BibitemOpen
		\bibfield  {author} {\bibinfo {author} {\bibfnamefont {E.}~\bibnamefont
				{Jeffrey}}, \bibinfo {author} {\bibfnamefont {D.}~\bibnamefont {Sank}},
			\bibinfo {author} {\bibfnamefont {J.}~\bibnamefont {Mutus}}, \bibinfo
			{author} {\bibfnamefont {T.}~\bibnamefont {White}}, \bibinfo {author}
			{\bibfnamefont {J.}~\bibnamefont {Kelly}}, \bibinfo {author} {\bibfnamefont
				{R.}~\bibnamefont {Barends}}, \bibinfo {author} {\bibfnamefont
				{Y.}~\bibnamefont {Chen}}, \bibinfo {author} {\bibfnamefont {Z.}~\bibnamefont
				{Chen}}, \bibinfo {author} {\bibfnamefont {B.}~\bibnamefont {Chiaro}},
			\bibinfo {author} {\bibfnamefont {A.}~\bibnamefont {Dunsworth}}, \emph
			{et~al.},\ }\href@noop {} {\bibfield  {journal} {\bibinfo  {journal} {Phys.
					Rev. Lett.}\ }\textbf {\bibinfo {volume} {112}},\ \bibinfo {pages} {190504}
			(\bibinfo {year} {2014})}\BibitemShut {NoStop}%
		\bibitem [{\citenamefont {Misra}\ and\ \citenamefont
			{Sudarshan}(1977)}]{misra1977zeno}%
		\BibitemOpen
		\bibfield  {author} {\bibinfo {author} {\bibfnamefont {B.}~\bibnamefont
				{Misra}}\ and\ \bibinfo {author} {\bibfnamefont {E.~G.}\ \bibnamefont
				{Sudarshan}},\ }\href@noop {} {\bibfield  {journal} {\bibinfo  {journal} {J.
					Math. Phys.}\ }\textbf {\bibinfo {volume} {18}},\ \bibinfo {pages} {756}
			(\bibinfo {year} {1977})}\BibitemShut {NoStop}%
		\bibitem [{\citenamefont {Itano}\ \emph {et~al.}(1990)\citenamefont {Itano},
			\citenamefont {Heinzen}, \citenamefont {Bollinger},\ and\ \citenamefont
			{Wineland}}]{itano1990quantum}%
		\BibitemOpen
		\bibfield  {author} {\bibinfo {author} {\bibfnamefont {W.~M.}\ \bibnamefont
				{Itano}}, \bibinfo {author} {\bibfnamefont {D.~J.}\ \bibnamefont {Heinzen}},
			\bibinfo {author} {\bibfnamefont {J.~J.}\ \bibnamefont {Bollinger}},\ and\
			\bibinfo {author} {\bibfnamefont {D.~J.}\ \bibnamefont {Wineland}},\
		}\href@noop {} {\bibfield  {journal} {\bibinfo  {journal} {Phys. Rev. A}\
			}\textbf {\bibinfo {volume} {41}},\ \bibinfo {pages} {2295} (\bibinfo {year}
			{1990})}\BibitemShut {NoStop}%
		\bibitem [{\citenamefont {Kofman}\ and\ \citenamefont
			{Kurizki}(2000)}]{kofman2000acceleration}%
		\BibitemOpen
		\bibfield  {author} {\bibinfo {author} {\bibfnamefont {A.}~\bibnamefont
				{Kofman}}\ and\ \bibinfo {author} {\bibfnamefont {G.}~\bibnamefont
				{Kurizki}},\ }\href@noop {} {\bibfield  {journal} {\bibinfo  {journal}
				{Nature}\ }\textbf {\bibinfo {volume} {405}},\ \bibinfo {pages} {546}
			(\bibinfo {year} {2000})}\BibitemShut {NoStop}%
		\bibitem [{\citenamefont {Kofman}\ and\ \citenamefont
			{Kurizki}(2001)}]{kofman2001frequent}%
		\BibitemOpen
		\bibfield  {author} {\bibinfo {author} {\bibfnamefont {A.}~\bibnamefont
				{Kofman}}\ and\ \bibinfo {author} {\bibfnamefont {G.}~\bibnamefont
				{Kurizki}},\ }\href@noop {} {\bibfield  {journal} {\bibinfo  {journal} {Z.
					Naturforsch A}\ }\textbf {\bibinfo {volume} {56}},\ \bibinfo {pages} {83}
			(\bibinfo {year} {2001})}\BibitemShut {NoStop}%
		\bibitem [{\citenamefont {Kofman}\ \emph {et~al.}(2001)\citenamefont {Kofman},
			\citenamefont {Kurizki},\ and\ \citenamefont {Opatrn{\`y}}}]{kofman2001zeno}%
		\BibitemOpen
		\bibfield  {author} {\bibinfo {author} {\bibfnamefont {A.}~\bibnamefont
				{Kofman}}, \bibinfo {author} {\bibfnamefont {G.}~\bibnamefont {Kurizki}},\
			and\ \bibinfo {author} {\bibfnamefont {T.}~\bibnamefont {Opatrn{\`y}}},\
		}\href@noop {} {\bibfield  {journal} {\bibinfo  {journal} {Phys. Rev. A}\
			}\textbf {\bibinfo {volume} {63}},\ \bibinfo {pages} {042108} (\bibinfo
			{year} {2001})}\BibitemShut {NoStop}%
		\bibitem [{\citenamefont {Kakuyanagi}\ \emph {et~al.}(2015)\citenamefont
			{Kakuyanagi}, \citenamefont {Baba}, \citenamefont {Matsuzaki}, \citenamefont
			{Nakano}, \citenamefont {Saito},\ and\ \citenamefont
			{Semba}}]{kakuyanagi2015observation}%
		\BibitemOpen
		\bibfield  {author} {\bibinfo {author} {\bibfnamefont {K.}~\bibnamefont
				{Kakuyanagi}}, \bibinfo {author} {\bibfnamefont {T.}~\bibnamefont {Baba}},
			\bibinfo {author} {\bibfnamefont {Y.}~\bibnamefont {Matsuzaki}}, \bibinfo
			{author} {\bibfnamefont {H.}~\bibnamefont {Nakano}}, \bibinfo {author}
			{\bibfnamefont {S.}~\bibnamefont {Saito}},\ and\ \bibinfo {author}
			{\bibfnamefont {K.}~\bibnamefont {Semba}},\ }\href@noop {} {\bibfield
			{journal} {\bibinfo  {journal} {New J. Phys.}\ }\textbf {\bibinfo {volume}
				{17}},\ \bibinfo {pages} {063035} (\bibinfo {year} {2015})}\BibitemShut
		{NoStop}%
		\bibitem [{\citenamefont {Slichter}\ \emph {et~al.}(2016)\citenamefont
			{Slichter}, \citenamefont {M{\"u}ller}, \citenamefont {Vijay}, \citenamefont
			{Weber}, \citenamefont {Blais},\ and\ \citenamefont
			{Siddiqi}}]{slichter2016quantum}%
		\BibitemOpen
		\bibfield  {author} {\bibinfo {author} {\bibfnamefont {D.}~\bibnamefont
				{Slichter}}, \bibinfo {author} {\bibfnamefont {C.}~\bibnamefont
				{M{\"u}ller}}, \bibinfo {author} {\bibfnamefont {R.}~\bibnamefont {Vijay}},
			\bibinfo {author} {\bibfnamefont {S.}~\bibnamefont {Weber}}, \bibinfo
			{author} {\bibfnamefont {A.}~\bibnamefont {Blais}},\ and\ \bibinfo {author}
			{\bibfnamefont {I.}~\bibnamefont {Siddiqi}},\ }\href@noop {} {\bibfield
			{journal} {\bibinfo  {journal} {New J. Phys.}\ }\textbf {\bibinfo {volume}
				{18}},\ \bibinfo {pages} {053031} (\bibinfo {year} {2016})}\BibitemShut
		{NoStop}%
		\bibitem [{\citenamefont {Hacohen-Gourgy}\ \emph {et~al.}(2018)\citenamefont
			{Hacohen-Gourgy}, \citenamefont {Garc{\'\i}a-Pintos}, \citenamefont {Martin},
			\citenamefont {Dressel},\ and\ \citenamefont
			{Siddiqi}}]{hacohen2018incoherent}%
		\BibitemOpen
		\bibfield  {author} {\bibinfo {author} {\bibfnamefont {S.}~\bibnamefont
				{Hacohen-Gourgy}}, \bibinfo {author} {\bibfnamefont {L.~P.}\ \bibnamefont
				{Garc{\'\i}a-Pintos}}, \bibinfo {author} {\bibfnamefont {L.~S.}\ \bibnamefont
				{Martin}}, \bibinfo {author} {\bibfnamefont {J.}~\bibnamefont {Dressel}},\
			and\ \bibinfo {author} {\bibfnamefont {I.}~\bibnamefont {Siddiqi}},\
		}\href@noop {} {\bibfield  {journal} {\bibinfo  {journal} {Phys. Rev. Lett.}\
			}\textbf {\bibinfo {volume} {120}},\ \bibinfo {pages} {020505} (\bibinfo
			{year} {2018})}\BibitemShut {NoStop}%
		\bibitem [{\citenamefont {Blumenthal}\ \emph {et~al.}(2022)\citenamefont
			{Blumenthal}, \citenamefont {Mor}, \citenamefont {Diringer}, \citenamefont
			{Martin}, \citenamefont {Lewalle}, \citenamefont {Burgarth}, \citenamefont
			{Whaley},\ and\ \citenamefont
			{Hacohen-Gourgy}}]{blumenthal2022demonstration}%
		\BibitemOpen
		\bibfield  {author} {\bibinfo {author} {\bibfnamefont {E.}~\bibnamefont
				{Blumenthal}}, \bibinfo {author} {\bibfnamefont {C.}~\bibnamefont {Mor}},
			\bibinfo {author} {\bibfnamefont {A.~A.}\ \bibnamefont {Diringer}}, \bibinfo
			{author} {\bibfnamefont {L.~S.}\ \bibnamefont {Martin}}, \bibinfo {author}
			{\bibfnamefont {P.}~\bibnamefont {Lewalle}}, \bibinfo {author} {\bibfnamefont
				{D.}~\bibnamefont {Burgarth}}, \bibinfo {author} {\bibfnamefont {K.~B.}\
				\bibnamefont {Whaley}},\ and\ \bibinfo {author} {\bibfnamefont
				{S.}~\bibnamefont {Hacohen-Gourgy}},\ }\href@noop {} {\bibfield  {journal}
			{\bibinfo  {journal} {npj Quantum Inf.}\ }\textbf {\bibinfo {volume} {8}},\
			\bibinfo {pages} {88} (\bibinfo {year} {2022})}\BibitemShut {NoStop}%
		\bibitem [{\citenamefont {Xiao}\ \emph {et~al.}(2023)\citenamefont {Xiao},
			\citenamefont {Thinga}, \citenamefont {Thorbeck}, \citenamefont {Govia},\
			and\ \citenamefont {Kamal}}]{DISCO}%
		\BibitemOpen
		\bibfield  {author} {\bibinfo {author} {\bibfnamefont {Z.}~\bibnamefont
				{Xiao}}, \bibinfo {author} {\bibfnamefont {J.}~\bibnamefont {Thinga}},
			\bibinfo {author} {\bibfnamefont {T.}~\bibnamefont {Thorbeck}}, \bibinfo
			{author} {\bibfnamefont {L.~C.~G.}\ \bibnamefont {Govia}},\ and\ \bibinfo
			{author} {\bibfnamefont {A.}~\bibnamefont {Kamal}}} (\bibinfo {year}
		{2023}),\ \bibinfo {note} {in preparation}\BibitemShut {NoStop}%
		\bibitem [{\citenamefont {Klimov}\ \emph {et~al.}(2018)\citenamefont {Klimov},
			\citenamefont {Kelly}, \citenamefont {Chen}, \citenamefont {Neeley},
			\citenamefont {Megrant}, \citenamefont {Burkett}, \citenamefont {Barends},
			\citenamefont {Arya}, \citenamefont {Chiaro}, \citenamefont {Chen} \emph
			{et~al.}}]{klimov2018fluctuations}%
		\BibitemOpen
		\bibfield  {author} {\bibinfo {author} {\bibfnamefont {P.}~\bibnamefont
				{Klimov}}, \bibinfo {author} {\bibfnamefont {J.}~\bibnamefont {Kelly}},
			\bibinfo {author} {\bibfnamefont {Z.}~\bibnamefont {Chen}}, \bibinfo {author}
			{\bibfnamefont {M.}~\bibnamefont {Neeley}}, \bibinfo {author} {\bibfnamefont
				{A.}~\bibnamefont {Megrant}}, \bibinfo {author} {\bibfnamefont
				{B.}~\bibnamefont {Burkett}}, \bibinfo {author} {\bibfnamefont
				{R.}~\bibnamefont {Barends}}, \bibinfo {author} {\bibfnamefont
				{K.}~\bibnamefont {Arya}}, \bibinfo {author} {\bibfnamefont {B.}~\bibnamefont
				{Chiaro}}, \bibinfo {author} {\bibfnamefont {Y.}~\bibnamefont {Chen}}, \emph
			{et~al.},\ }\href@noop {} {\bibfield  {journal} {\bibinfo  {journal} {Phys.
					Rev. Lett.}\ }\textbf {\bibinfo {volume} {121}},\ \bibinfo {pages} {090502}
			(\bibinfo {year} {2018})}\BibitemShut {NoStop}%
		\bibitem [{\citenamefont {Carroll}\ \emph {et~al.}(2022)\citenamefont
			{Carroll}, \citenamefont {Rosenblatt}, \citenamefont {Jurcevic},
			\citenamefont {Lauer},\ and\ \citenamefont {Kandala}}]{carroll2022dynamics}%
		\BibitemOpen
		\bibfield  {author} {\bibinfo {author} {\bibfnamefont {M.}~\bibnamefont
				{Carroll}}, \bibinfo {author} {\bibfnamefont {S.}~\bibnamefont {Rosenblatt}},
			\bibinfo {author} {\bibfnamefont {P.}~\bibnamefont {Jurcevic}}, \bibinfo
			{author} {\bibfnamefont {I.}~\bibnamefont {Lauer}},\ and\ \bibinfo {author}
			{\bibfnamefont {A.}~\bibnamefont {Kandala}},\ }\href@noop {} {\bibfield
			{journal} {\bibinfo  {journal} {npj Quantum Inf.}\ }\textbf {\bibinfo
				{volume} {8}},\ \bibinfo {pages} {132} (\bibinfo {year} {2022})}\BibitemShut
		{NoStop}%
		\bibitem [{\citenamefont {Barends}\ \emph {et~al.}(2013)\citenamefont
			{Barends}, \citenamefont {Kelly}, \citenamefont {Megrant}, \citenamefont
			{Sank}, \citenamefont {Jeffrey}, \citenamefont {Chen}, \citenamefont {Yin},
			\citenamefont {Chiaro}, \citenamefont {Mutus}, \citenamefont {Neill} \emph
			{et~al.}}]{barends2013coherent}%
		\BibitemOpen
		\bibfield  {author} {\bibinfo {author} {\bibfnamefont {R.}~\bibnamefont
				{Barends}}, \bibinfo {author} {\bibfnamefont {J.}~\bibnamefont {Kelly}},
			\bibinfo {author} {\bibfnamefont {A.}~\bibnamefont {Megrant}}, \bibinfo
			{author} {\bibfnamefont {D.}~\bibnamefont {Sank}}, \bibinfo {author}
			{\bibfnamefont {E.}~\bibnamefont {Jeffrey}}, \bibinfo {author} {\bibfnamefont
				{Y.}~\bibnamefont {Chen}}, \bibinfo {author} {\bibfnamefont {Y.}~\bibnamefont
				{Yin}}, \bibinfo {author} {\bibfnamefont {B.}~\bibnamefont {Chiaro}},
			\bibinfo {author} {\bibfnamefont {J.}~\bibnamefont {Mutus}}, \bibinfo
			{author} {\bibfnamefont {C.}~\bibnamefont {Neill}}, \emph {et~al.},\
		}\href@noop {} {\bibfield  {journal} {\bibinfo  {journal} {Phys. Rev. Lett.}\
			}\textbf {\bibinfo {volume} {111}},\ \bibinfo {pages} {080502} (\bibinfo
			{year} {2013})}\BibitemShut {NoStop}%
		\bibitem [{\citenamefont {Lisenfeld}\ \emph {et~al.}(2023)\citenamefont
			{Lisenfeld}, \citenamefont {Bilmes},\ and\ \citenamefont
			{Ustinov}}]{lisenfeld2023enhancing}%
		\BibitemOpen
		\bibfield  {author} {\bibinfo {author} {\bibfnamefont {J.}~\bibnamefont
				{Lisenfeld}}, \bibinfo {author} {\bibfnamefont {A.}~\bibnamefont {Bilmes}},\
			and\ \bibinfo {author} {\bibfnamefont {A.~V.}\ \bibnamefont {Ustinov}},\
		}\href@noop {} {\bibfield  {journal} {\bibinfo  {journal} {npj Quantum Inf.}\
			}\textbf {\bibinfo {volume} {9}},\ \bibinfo {pages} {8} (\bibinfo {year}
			{2023})}\BibitemShut {NoStop}%
		\bibitem [{\citenamefont {M{\"u}ller}\ \emph {et~al.}(2019)\citenamefont
			{M{\"u}ller}, \citenamefont {Cole},\ and\ \citenamefont
			{Lisenfeld}}]{muller2019towards}%
		\BibitemOpen
		\bibfield  {author} {\bibinfo {author} {\bibfnamefont {C.}~\bibnamefont
				{M{\"u}ller}}, \bibinfo {author} {\bibfnamefont {J.~H.}\ \bibnamefont
				{Cole}},\ and\ \bibinfo {author} {\bibfnamefont {J.}~\bibnamefont
				{Lisenfeld}},\ }\href@noop {} {\bibfield  {journal} {\bibinfo  {journal}
				{Rep.Prog. Phys.}\ }\textbf {\bibinfo {volume} {82}},\ \bibinfo {pages}
			{124501} (\bibinfo {year} {2019})}\BibitemShut {NoStop}%
		\bibitem [{sup()}]{supplement}%
		\BibitemOpen
		\bibinfo {title} {See supplementary information}\BibitemShut {NoStop}%
		\bibitem [{\citenamefont {Gambetta}\ \emph {et~al.}(2006)\citenamefont
			{Gambetta}, \citenamefont {Blais}, \citenamefont {Schuster}, \citenamefont
			{Wallraff}, \citenamefont {Frunzio}, \citenamefont {Majer}, \citenamefont
			{Devoret}, \citenamefont {Girvin},\ and\ \citenamefont
			{Schoelkopf}}]{gambetta2006qubit}%
		\BibitemOpen
		\bibfield  {title} {  }\bibfield  {author} {\bibinfo {author} {\bibfnamefont
				{J.}~\bibnamefont {Gambetta}}, \bibinfo {author} {\bibfnamefont
				{A.}~\bibnamefont {Blais}}, \bibinfo {author} {\bibfnamefont {D.~I.}\
				\bibnamefont {Schuster}}, \bibinfo {author} {\bibfnamefont {A.}~\bibnamefont
				{Wallraff}}, \bibinfo {author} {\bibfnamefont {L.}~\bibnamefont {Frunzio}},
			\bibinfo {author} {\bibfnamefont {J.}~\bibnamefont {Majer}}, \bibinfo
			{author} {\bibfnamefont {M.~H.}\ \bibnamefont {Devoret}}, \bibinfo {author}
			{\bibfnamefont {S.~M.}\ \bibnamefont {Girvin}},\ and\ \bibinfo {author}
			{\bibfnamefont {R.~J.}\ \bibnamefont {Schoelkopf}},\ }\href@noop {}
		{\bibfield  {journal} {\bibinfo  {journal} {Phys. Rev. A}\ }\textbf {\bibinfo
				{volume} {74}},\ \bibinfo {pages} {042318} (\bibinfo {year}
			{2006})}\BibitemShut {NoStop}%
		\bibitem [{\citenamefont {Thorbeck}\ \emph {et~al.}(2022)\citenamefont
			{Thorbeck}, \citenamefont {Eddins}, \citenamefont {Lauer}, \citenamefont
			{McClure},\ and\ \citenamefont {Carroll}}]{thorbeck2022tls}%
		\BibitemOpen
		\bibfield  {author} {\bibinfo {author} {\bibfnamefont {T.}~\bibnamefont
				{Thorbeck}}, \bibinfo {author} {\bibfnamefont {A.}~\bibnamefont {Eddins}},
			\bibinfo {author} {\bibfnamefont {I.}~\bibnamefont {Lauer}}, \bibinfo
			{author} {\bibfnamefont {D.~T.}\ \bibnamefont {McClure}},\ and\ \bibinfo
			{author} {\bibfnamefont {M.}~\bibnamefont {Carroll}},\ }\href@noop {}
		{\bibfield  {journal} {\bibinfo  {journal} {arXiv:2210.04780}\ } (\bibinfo
			{year} {2022})}\BibitemShut {NoStop}%
		\bibitem [{\citenamefont {Ai}\ \emph {et~al.}(2013)\citenamefont {Ai},
			\citenamefont {Xu}, \citenamefont {Yi}, \citenamefont {Kofman}, \citenamefont
			{Sun},\ and\ \citenamefont {Nori}}]{ai2013quantum}%
		\BibitemOpen
		\bibfield  {author} {\bibinfo {author} {\bibfnamefont {Q.}~\bibnamefont
				{Ai}}, \bibinfo {author} {\bibfnamefont {D.}~\bibnamefont {Xu}}, \bibinfo
			{author} {\bibfnamefont {S.}~\bibnamefont {Yi}}, \bibinfo {author}
			{\bibfnamefont {A.}~\bibnamefont {Kofman}}, \bibinfo {author} {\bibfnamefont
				{C.}~\bibnamefont {Sun}},\ and\ \bibinfo {author} {\bibfnamefont
				{F.}~\bibnamefont {Nori}},\ }\href@noop {} {\bibfield  {journal} {\bibinfo
				{journal} {Sci. Rep.}\ }\textbf {\bibinfo {volume} {3}},\ \bibinfo {pages}
			{1} (\bibinfo {year} {2013})}\BibitemShut {NoStop}%
		\bibitem [{\citenamefont {Alvarez}\ \emph {et~al.}(2010)\citenamefont
			{Alvarez}, \citenamefont {Rao}, \citenamefont {Frydman},\ and\ \citenamefont
			{Kurizki}}]{alvarez2010zeno}%
		\BibitemOpen
		\bibfield  {author} {\bibinfo {author} {\bibfnamefont {G.~A.}\ \bibnamefont
				{Alvarez}}, \bibinfo {author} {\bibfnamefont {D.~B.}\ \bibnamefont {Rao}},
			\bibinfo {author} {\bibfnamefont {L.}~\bibnamefont {Frydman}},\ and\ \bibinfo
			{author} {\bibfnamefont {G.}~\bibnamefont {Kurizki}},\ }\href@noop {}
		{\bibfield  {journal} {\bibinfo  {journal} {Phys. Rev. Lett.}\ }\textbf
			{\bibinfo {volume} {105}},\ \bibinfo {pages} {160401} (\bibinfo {year}
			{2010})}\BibitemShut {NoStop}%
		\bibitem [{\citenamefont {Cao}\ \emph {et~al.}(2012)\citenamefont {Cao},
			\citenamefont {Ai}, \citenamefont {Sun},\ and\ \citenamefont
			{Nori}}]{cao2012transition}%
		\BibitemOpen
		\bibfield  {author} {\bibinfo {author} {\bibfnamefont {X.}~\bibnamefont
				{Cao}}, \bibinfo {author} {\bibfnamefont {Q.}~\bibnamefont {Ai}}, \bibinfo
			{author} {\bibfnamefont {C.-P.}\ \bibnamefont {Sun}},\ and\ \bibinfo {author}
			{\bibfnamefont {F.}~\bibnamefont {Nori}},\ }\href@noop {} {\bibfield
			{journal} {\bibinfo  {journal} {Phys. Lett. A}\ }\textbf {\bibinfo {volume}
				{376}},\ \bibinfo {pages} {349} (\bibinfo {year} {2012})}\BibitemShut
		{NoStop}%
		\bibitem [{\citenamefont {Auff{\`e}ves}\ \emph {et~al.}(2010)\citenamefont
			{Auff{\`e}ves}, \citenamefont {Gerace}, \citenamefont {G{\'e}rard},
			\citenamefont {Santos}, \citenamefont {Andreani},\ and\ \citenamefont
			{Poizat}}]{auffeves2010controlling}%
		\BibitemOpen
		\bibfield  {author} {\bibinfo {author} {\bibfnamefont {A.}~\bibnamefont
				{Auff{\`e}ves}}, \bibinfo {author} {\bibfnamefont {D.}~\bibnamefont
				{Gerace}}, \bibinfo {author} {\bibfnamefont {J.-M.}\ \bibnamefont
				{G{\'e}rard}}, \bibinfo {author} {\bibfnamefont {M.~F.}\ \bibnamefont
				{Santos}}, \bibinfo {author} {\bibfnamefont {L.}~\bibnamefont {Andreani}},\
			and\ \bibinfo {author} {\bibfnamefont {J.-P.}\ \bibnamefont {Poizat}},\
		}\href@noop {} {\bibfield  {journal} {\bibinfo  {journal} {Phys. Rev. B}\
			}\textbf {\bibinfo {volume} {81}},\ \bibinfo {pages} {245419} (\bibinfo
			{year} {2010})}\BibitemShut {NoStop}%
		\bibitem [{\citenamefont {Clerk}\ \emph {et~al.}(2010)\citenamefont {Clerk},
			\citenamefont {Devoret}, \citenamefont {Girvin}, \citenamefont {Marquardt},\
			and\ \citenamefont {Schoelkopf}}]{clerk2010introduction}%
		\BibitemOpen
		\bibfield  {author} {\bibinfo {author} {\bibfnamefont {A.~A.}\ \bibnamefont
				{Clerk}}, \bibinfo {author} {\bibfnamefont {M.~H.}\ \bibnamefont {Devoret}},
			\bibinfo {author} {\bibfnamefont {S.~M.}\ \bibnamefont {Girvin}}, \bibinfo
			{author} {\bibfnamefont {F.}~\bibnamefont {Marquardt}},\ and\ \bibinfo
			{author} {\bibfnamefont {R.~J.}\ \bibnamefont {Schoelkopf}},\ }\href@noop {}
		{\bibfield  {journal} {\bibinfo  {journal} {Rev. Mod. Phys.}\ }\textbf
			{\bibinfo {volume} {82}},\ \bibinfo {pages} {1155} (\bibinfo {year}
			{2010})}\BibitemShut {NoStop}%
		\bibitem [{\citenamefont {Girvin}(2014)}]{girvin2014circuit}%
		\BibitemOpen
		\bibfield  {author} {\bibinfo {author} {\bibfnamefont {S.~M.}\ \bibnamefont
				{Girvin}},\ }\bibinfo {title} {Circuit {QED}: superconducting qubits coupled
			to microwave photons},\ in\ \href@noop {} {\emph {\bibinfo {booktitle}
				{Quantum machines: measurement and control of engineered quantum systems}}}\
		(\bibinfo {year} {2014})\ pp.\ \bibinfo {pages} {113--256}\BibitemShut
		{NoStop}%
		\bibitem [{\citenamefont {Schulman}(1998)}]{schulman1998continuous}%
		\BibitemOpen
		\bibfield  {author} {\bibinfo {author} {\bibfnamefont {L.}~\bibnamefont
				{Schulman}},\ }\href@noop {} {\bibfield  {journal} {\bibinfo  {journal}
				{Phys. Rev. A}\ }\textbf {\bibinfo {volume} {57}},\ \bibinfo {pages} {1509}
			(\bibinfo {year} {1998})}\BibitemShut {NoStop}%
		\bibitem [{\citenamefont {Lidar}(2019)}]{Lidar}%
		\BibitemOpen
		\bibfield  {author} {\bibinfo {author} {\bibfnamefont {D.~A.}\ \bibnamefont
				{Lidar}},\ }\href@noop {} {} (\bibinfo {year} {2019}),\ \Eprint
		{https://arxiv.org/abs/arXiv:1902.00967} {arXiv:1902.00967} \BibitemShut
		{NoStop}%
	\end{thebibliography}
\end{document}